\documentclass[default,iicol]{sn-jnl}% Default with double column layout

\usepackage{flushend}

\jyear{2023}%

\theoremstyle{thmstyleone}%
\theoremstyle{thmstyletwo}%

\theoremstyle{thmstylethree}%

\begin{document}

%\title[Niel's Chess]{Niel's Chess\textsuperscript{\textsuperscript{\textregistered}}\,- The Battle of the Quantum Age}
\title[Niel's Chess]{Niel's Chess\texttrademark\,\,- The Battle of the Quantum Age}

\author*{\fnm{Tamás} \sur{Varga}}\email{tvarga@q-edu-lab.com}
\affil{\orgname{q-edu-lab.com}, \orgaddress{\state{Zurich}, \country{Switzerland}}}

%\author*{\fnm{Tamás} \sur{Varga}}\email{varga78t@gmail.com}
%\affil{\orgaddress{\city{Männedorf}, \state{Zurich}, \country{Switzerland}}}

\abstract{In this paper, a quantum variant of chess is introduced, which can be played on a traditional board without the need of using computers or other electronic devices. The rules of the game arise naturally by combining the rules of conventional chess with key quantum-physical effects such as superposition and entanglement. Niel's Chess is recommended for ages 10 and above, to everyone who wishes to play a creative game with historical roots and at the same time gain intuition about the foundational quantum effects that power cutting-edge technologies like quantum computing and quantum communication, which are poised to revolutionise our society in the coming decades.}

\keywords{Niel's Chess, quantum chess, superposition, entanglement, traditional board game}

\maketitle

%\pagestyle{myheadings}
%\markboth{\textcolor{red}{CONFIDENTIAL - DO NOT DISTRIBUTE}}{\textcolor{red}{CONFIDENTIAL - DO NOT DISTRIBUTE}}

\section{Background}\label{background}

Quantum information technology is on the cusp of disrupting a wide range of industries \cite{mckinsey}. In the next decades, quantum computers are expected to help solve certain problems in mere hours or days that would otherwise take billions of years on conventional computers. Moreover, using quantum-physical effects makes it possible to devise communication protocols whose security is guaranteed by the laws of nature, as opposed to relying on mathematical problems that are \textit{believed} to be hard to solve.

At the heart of the upcoming quantum revolution lies the quantum effect \textit{superposition}, which means that a physical system is in a combination of two or more states that are mutually exclusive. For example, according to quantum theory, an atom's location can be a combination of two or more different locations, somehow as if the atom was in multiple places at the same time. However, when someone tries to observe the atom to see what it looks like being at two or more locations at once, the superposition immediately 'collapses' and the atom will be found at exactly one of those locations, picked by nature in a truly random fashion; nobody in the universe can predict the outcome with certainty. What's more, superposition can also lead to the even stranger quantum effect \textit{entanglement}, where two or more distinct, typically spatially separated physical systems seem to be connected despite having no apparent physical link between them. It's as if two coins tossed in two far-away cities were destined to (randomly) land on the same side.

As quantum information technologies become more and more widespread in the future, the aforementioned quantum effects will gradually penetrate into our professional as well as personal lives. Niel's Chess can be an example of the latter, a traditional board game in which the key to success is a better acquaintance with superposition and entanglement.

Besides being an interesting logic game, Niel's Chess can also serve educational purposes, helping educators prepare the wide public, especially the younger generations, for the quantum era our society is heading towards.

\section{Previous works}\label{previous}

Previous works to extend chess to the quantum realm include, most notably, \cite{akl} and \cite{cantwell}.

In \cite{akl}, the idea is that each piece on the board, except the king, can be in a superposition state, being of two types at once. E.g. a piece may be a knight and a bishop at once, and when the player touches the piece it 'collapses' with equal probability to be either a knight or a bishop, and then the move can be made with it accordingly. Or, if there is no possible move, the player's turn is over. If a piece lands on a black square, it regains its original superposition state, otherwise it remains in a conventional (i.e. collapsed) state. Every piece (except the king) starts out in a random superposition state, but initially neither of the players know in what superposition any given piece exactly is; they gain information about that only by seeing the piece collapse. In the most extreme case, a piece might be in a superposition of being a pawn and a queen. The king may be placed, or left, in check, and the game ends when a player captures the opponent's king.

In \cite{cantwell}, a different philosophy is used. Instead of the individual pieces, it is the whole game that is in a superposition state. It's a superposition of conventional chessboards, each having a different position of the pieces on it, as if multiple related chess games were being played in parallel. A computer keeps track of the superposition state of the game, and there is a specific (partial) collapse rule, executed by the system whenever necessary before a move is made, which uses a scheme to (randomly) remove boards from the superposition, so as to prevent the overall situation from becoming unmanageable in terms of execution and visual representation. There is no notion of check or checkmate, and similarly to \cite{akl} the king can be captured like any other piece. The game ends as soon as a player doesn't have a king on any of the boards in the superposition.

\section{Design principles}\label{principles}

The quantum-chess variant presented here takes a different approach compared to \cite{akl} and \cite{cantwell}. (See Appendix F for a summary.)

Here is a list of the most important guiding principles:

\begin{enumerate}[1.]
\item \textbf{Traditional: }the game should be played without relying on computers or other electronic devices.
\item \textbf{It's chess: }preserve the characteristics of chess pieces, their movements, the notion of check, and the visibility of the game's state for both players.
\item \textbf{Logic game: }the quantum world is inherently random, but luck shouldn't make up for lack of skill.
\item \textbf{Intuitive: }make quantum rules arise naturally by combining quantum effects with the rules of chess.
\end{enumerate}

As a result, no quantum effect is introduced just for the sake of it. All quantum rules are simple and harmonise with the characteristics and spirit of chess. One crucial example is that the king cannot be captured, only checkmated. Furthermore, the complete state of the game can be seen by looking at the chessboard. Nothing is hidden, or difficult to calculate. The addition of quantum rules enriches chess with creative ideas in both attack and defence.

To lower the barrier to entry for newcomers, it is recommended to start by playing the game on smaller boards and/or with simplified rules, see Appendix E and Appendix A for suggestions, respectively.

\section{Basic rules of play - Part 1}\label{basic1}

The "Basic Rules of Play" section of the FIDE Laws of Chess document \cite{fide} is taken as baseline, in which Articles 1 to 5 describe the non-competitive rules of chess.

The five FIDE Laws of Chess articles are extended here by the additional Articles 6 to 8 specifying the non-competitive quantum rules of Niel's Chess.

\subsection*{Article 6: Indefinite pieces}

\begin{enumerate}[1.1]
\item[6.1] For every conventional piece there is a corresponding pair of 'indefinite pieces', including one piece with a red mark on it and another with blue.
\begin{figure}[H]
\centering
\includegraphics[width=0.9\columnwidth]{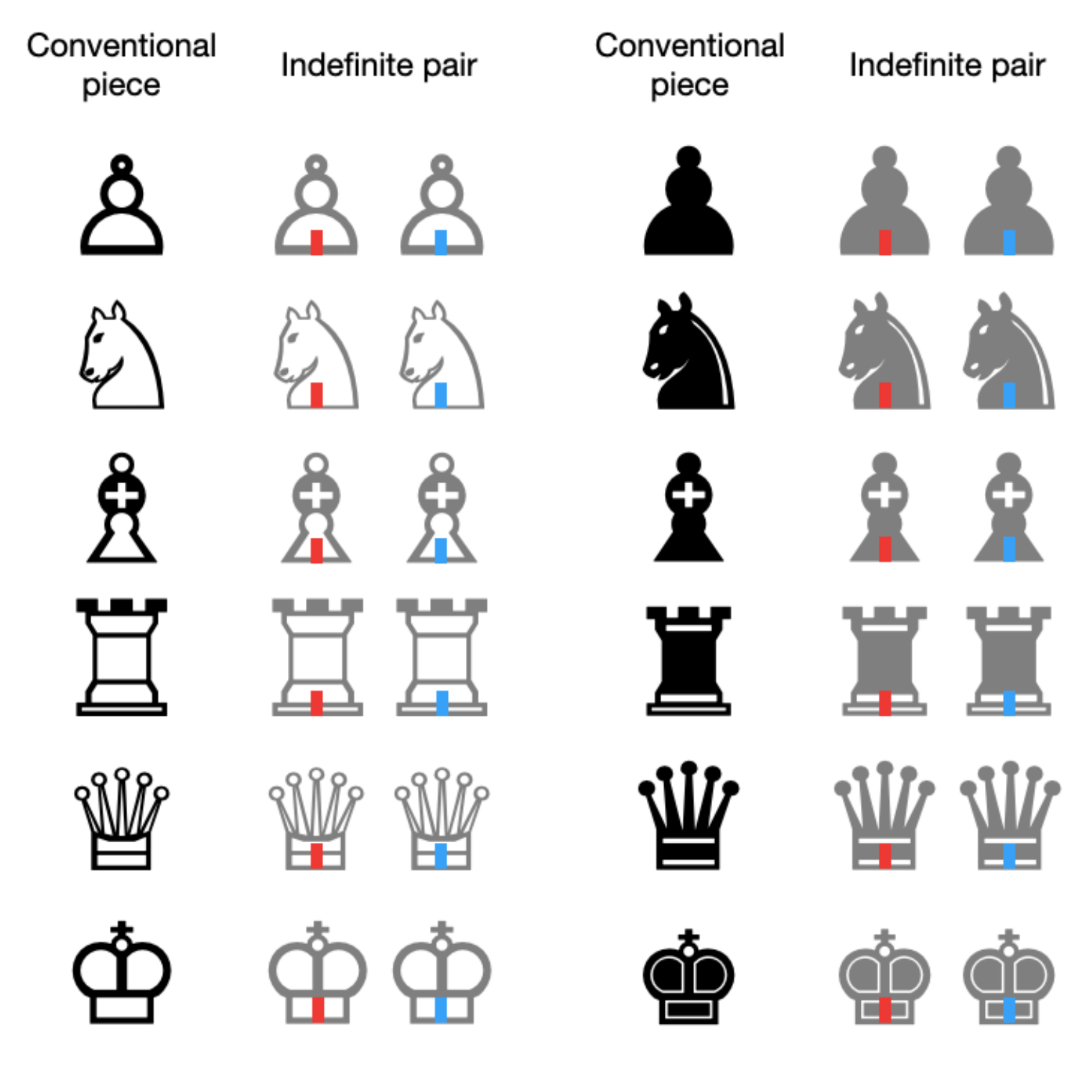}
\end{figure}
\item[6.2] Articles 1 to 5 apply to situations involving indefinite pieces, unless specified otherwise in Articles 7 and 8. Articles 6.2.1 to 6.2.6 stand here only to help understanding:\\
\begin{enumerate}[1.1.1]
\item[6.2.1] \textit{[Clarification]} In accordance with Article 1.4.1, leaving one’s own indefinite king under attack, exposing the same to attack and also capturing the opponent’s indefinite king is not allowed, barring Articles 8.8 and 8.9.
\item[6.2.2] \textit{[Clarification]} An indefinite piece may move to a square in accordance with the rules for conventional pieces stipulated in Articles 3.1 to 3.9, with the exception that the capturing part and the pawn-promotion part are governed by Articles 7 and 8. It's not allowed to move over indefinite intervening pieces.
\item[6.2.3] \textit{[Clarification]} In accordance with Article 3.1.2, a piece, indefinite or not, is said to attack an opponent’s piece, indefinite or not, if the piece could make (or at least attempt, see Article 7.4) a capture on that square according to Articles 3.2 to 3.8 (disregarding any other rule).
\item[6.2.4] \textit{[Clarification]} In accordance with Article 3.8.2.1, castling is not allowed with an indefinite king or rook.
\item[6.2.5] \textit{[Clarification]} In accordance with Article 3.8.2.2, castling is prevented temporarily if the square on which the king stands, or the square which it must cross, or the square which it is to occupy, is attacked by one or more of the opponent's indefinite pieces.
\item[6.2.6] \textit{[Clarification]} In accordance with Article 2.3, initially there are no indefinite pieces on the chessboard.
\begin{figure}[H]
\centering
\includegraphics[width=0.9\columnwidth]{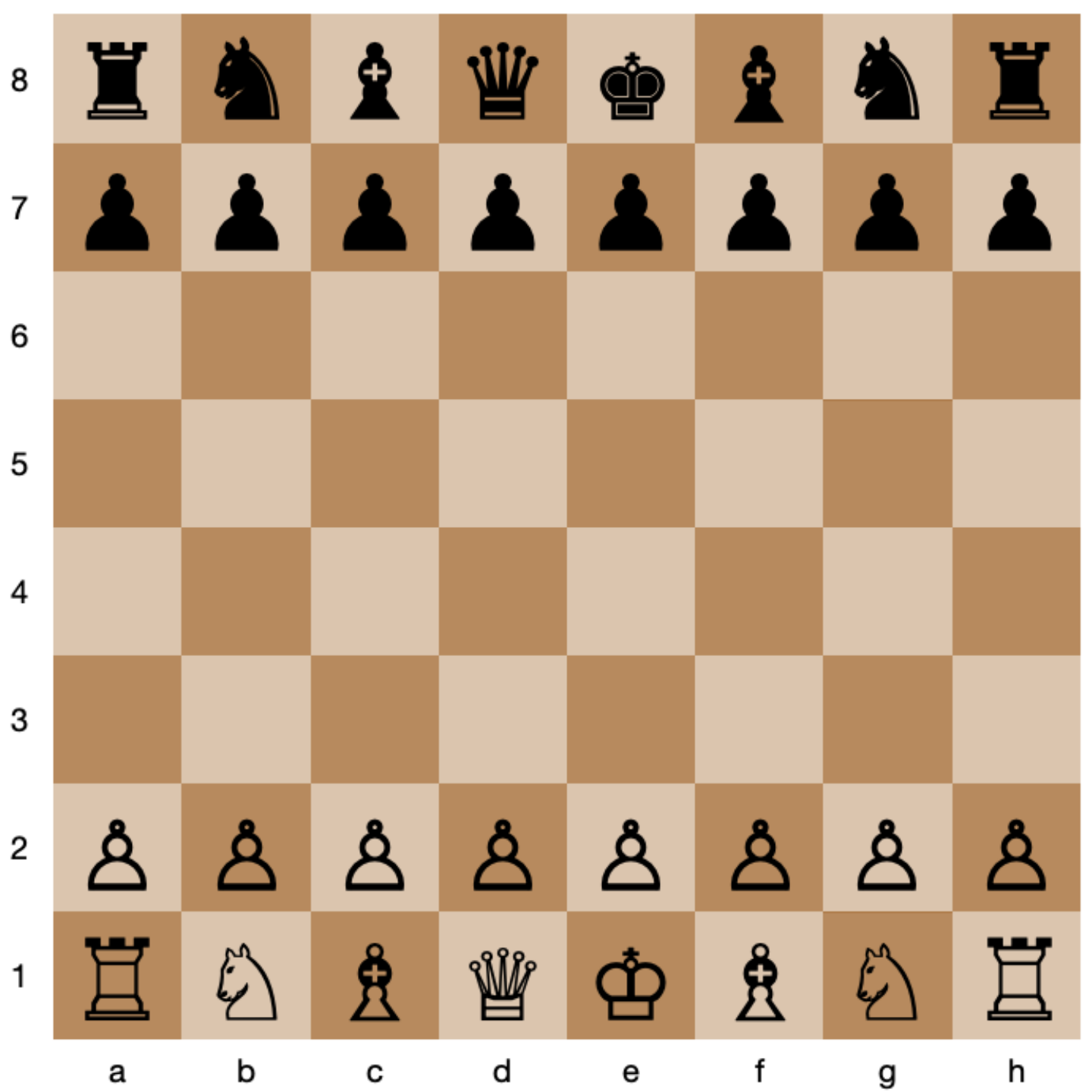}
\end{figure}
\end{enumerate}
\item[6.3] Except for indefinite pawns, all instances of indefinite pairs of the same piece and colour must have their marks with different fill patterns. (This is to avoid ambiguity as to which indefinite pieces on the chessboard are paired, see Article 7.1.)
\begin{figure}[H]
\centering
\includegraphics[width=0.9\columnwidth]{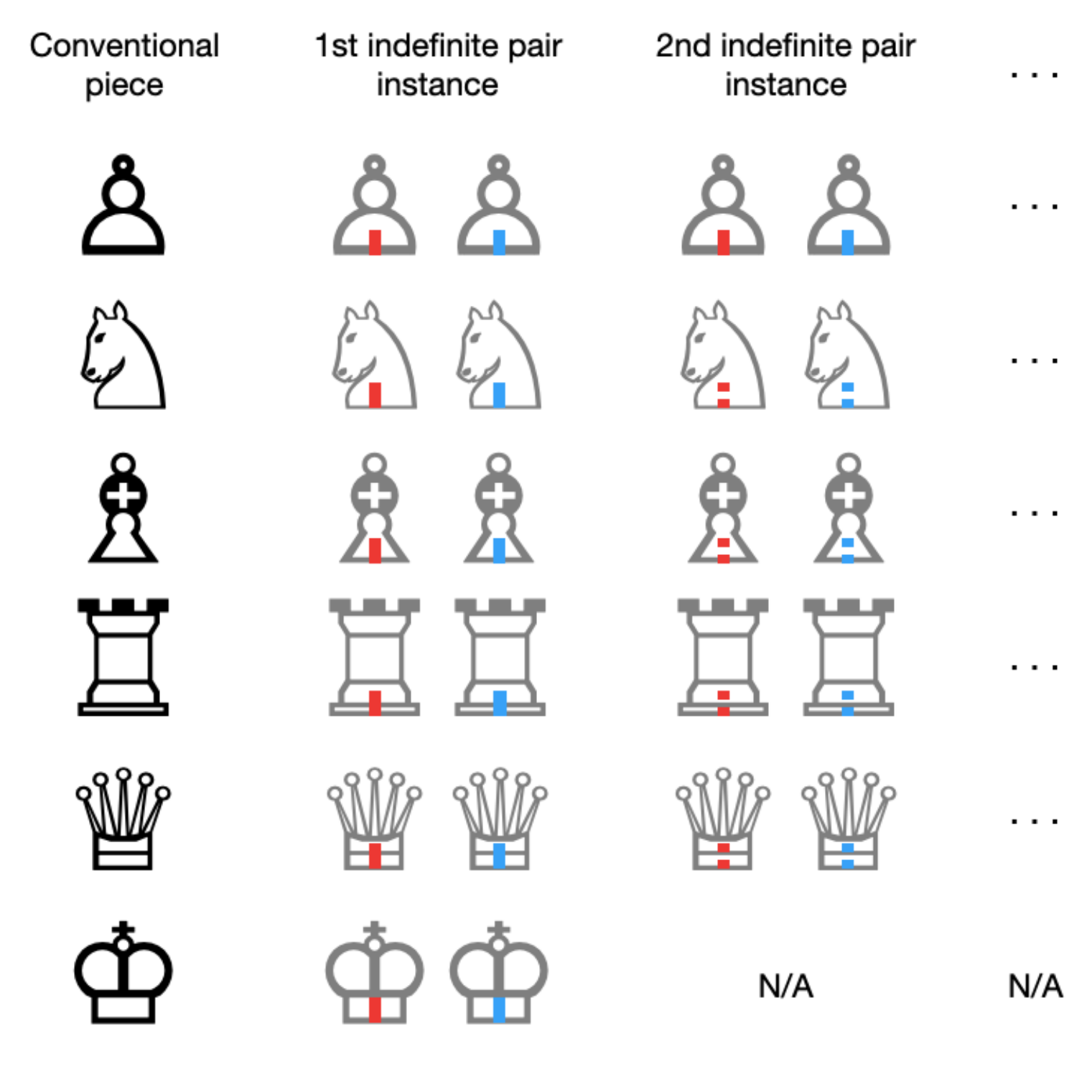}
\end{figure}
\begin{figure}[H]
\centering
\includegraphics[width=0.9\columnwidth]{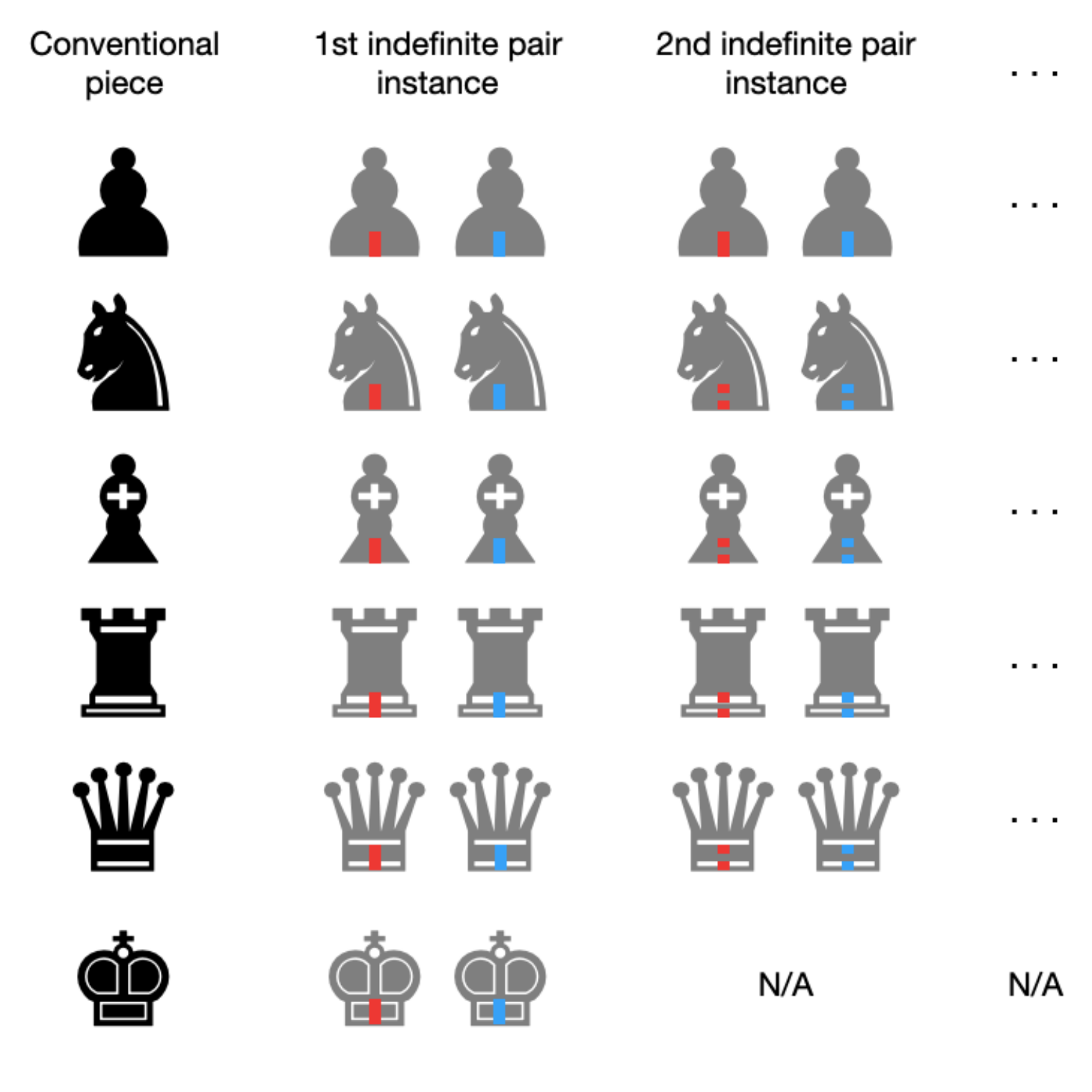}
\end{figure}
\end{enumerate}

\subsection*{Article 7: Superposition rules}

\begin{enumerate}[1.1]
\item[7.1] A conventional piece may move to two unoccupied squares at once (each reachable as per Articles 3.1 to 3.8), or move to one unoccupied square and stay where it is at once. This is called a 'superposition move'. The conventional piece has to be replaced by the two pieces of a corresponding indefinite pair instance, placed on the two squares of arrival, one each. The two indefinite pieces are said to be 'paired' (as they represent the conventional piece in a superposition state of being on two squares at once).
\begin{figure}[H]
\centering
\includegraphics[width=0.9\columnwidth]{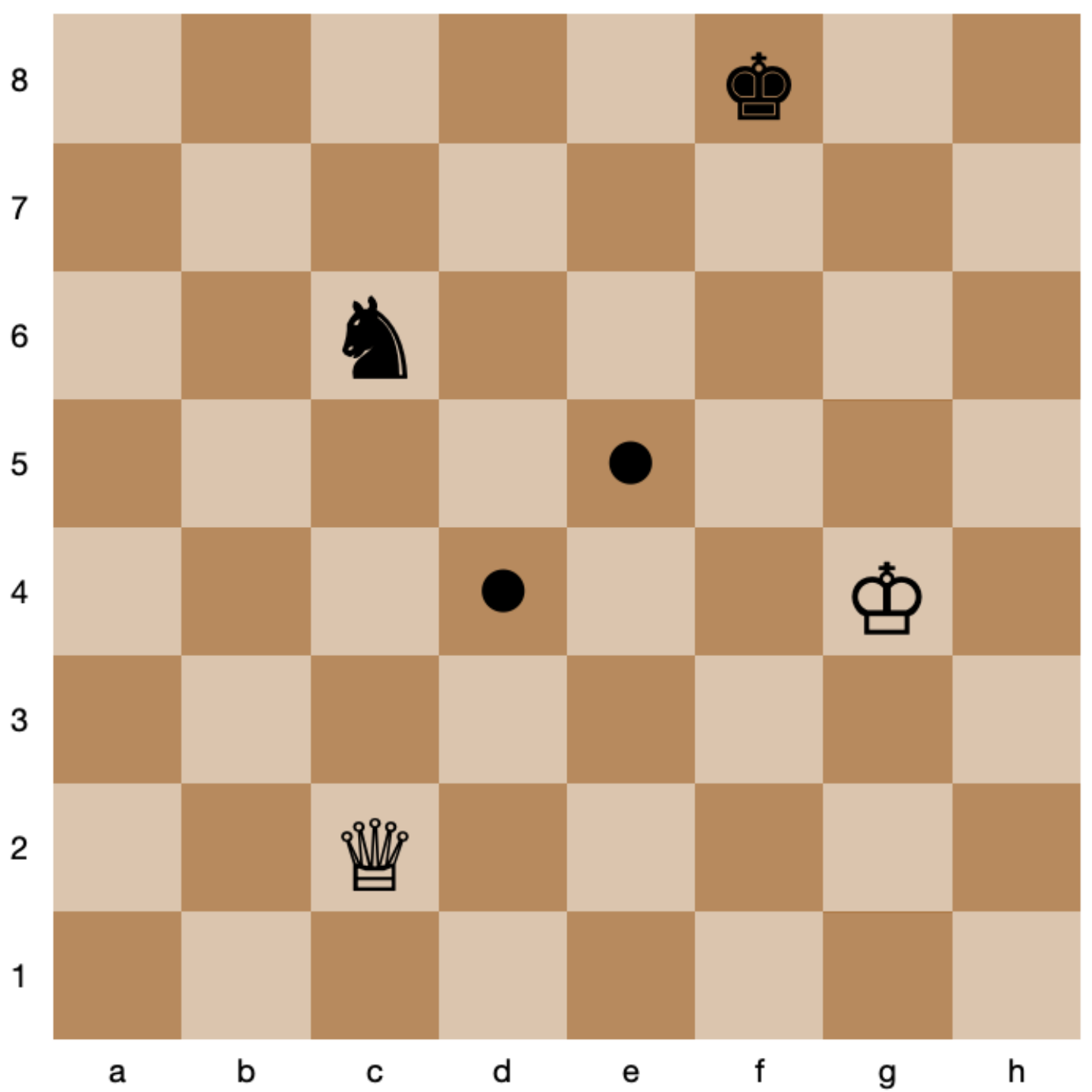}
\end{figure}
\item[7.2] If the marks on the paired indefinite pieces both face the opponent, or both face the player who owns the pair, it is said to be an 'equal superposition' (using opponent-facing marks are recommended for an easier overview). If one mark faces the opponent and the other the player who owns the pair, it is called an 'unequal superposition'. The player making the superposition move can freely choose either option. No other possibilities are allowed (i.e. only facing the upper or lower side of the square is permitted).\footnote{For better visibility in 2D, the red and blue marks are displayed separated from their respective pieces.}
\begin{figure}[H]
\centering
\includegraphics[width=0.9\columnwidth]{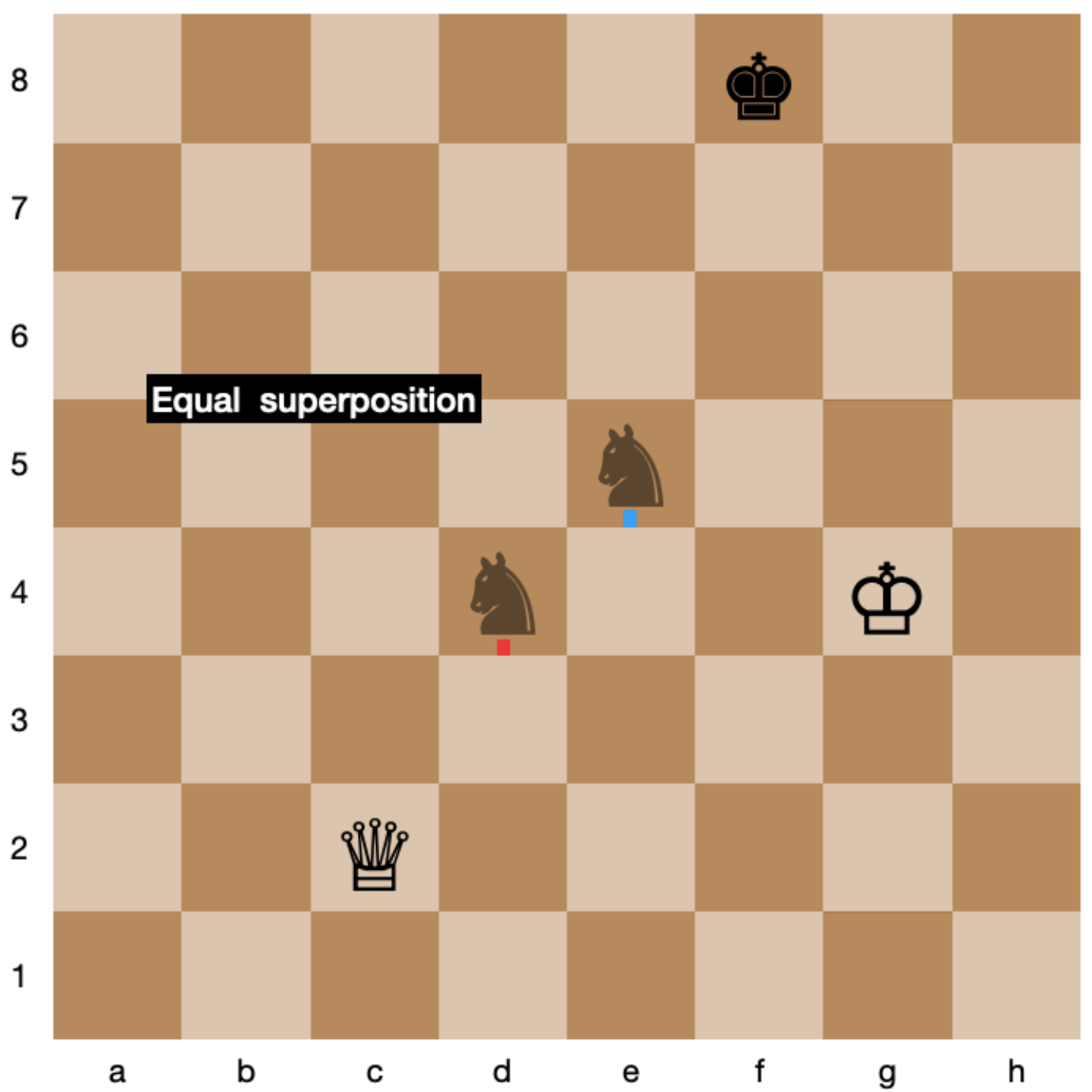}
\vspace*{-0.5cm}
\end{figure}
\begin{figure}[H]
\centering
\includegraphics[width=0.9\columnwidth]{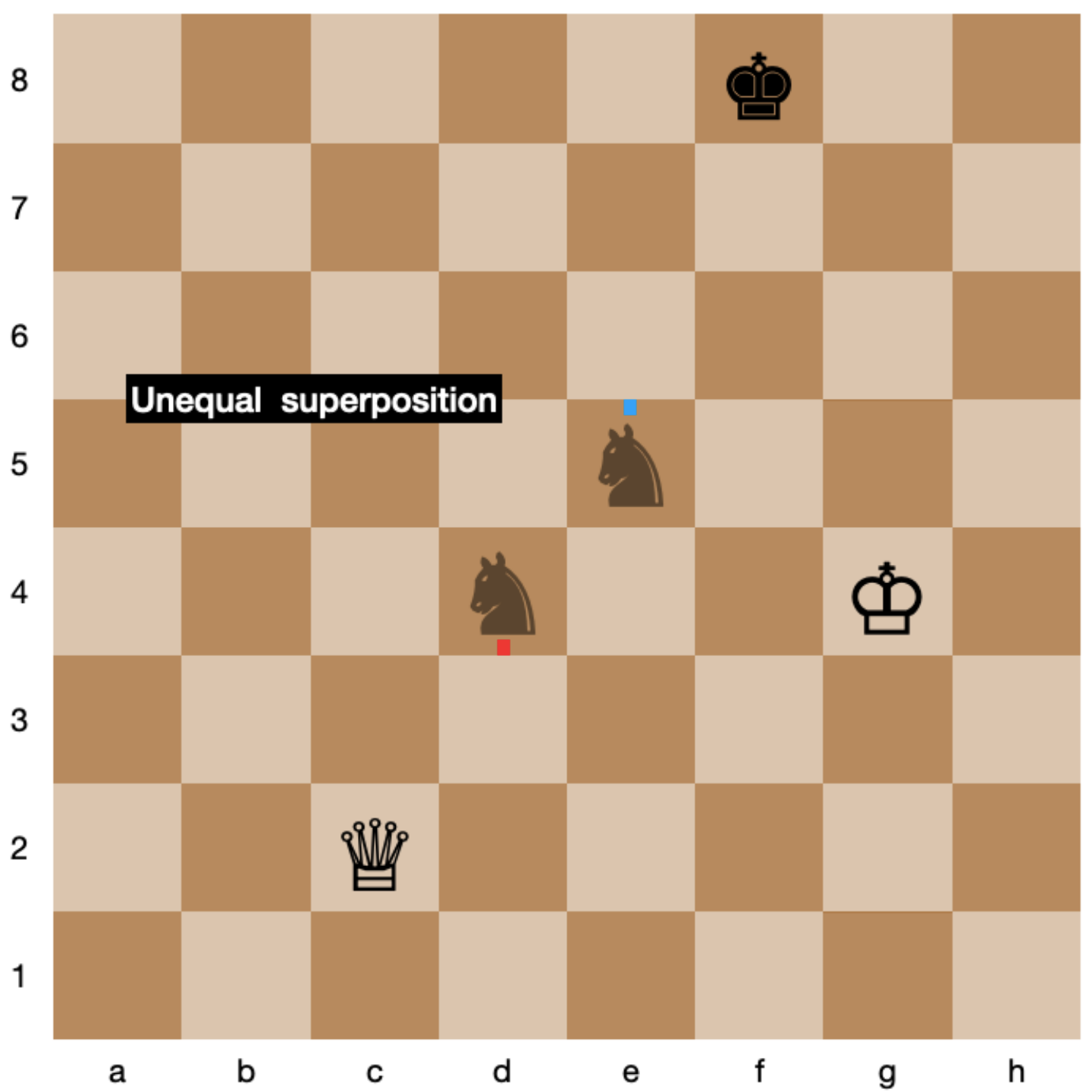}
\end{figure}
\item[7.3] For paired indefinite pawns, the piece on the rank farther from its starting position must have the red mark.
\begin{figure}[H]
\centering
\includegraphics[width=0.9\columnwidth]{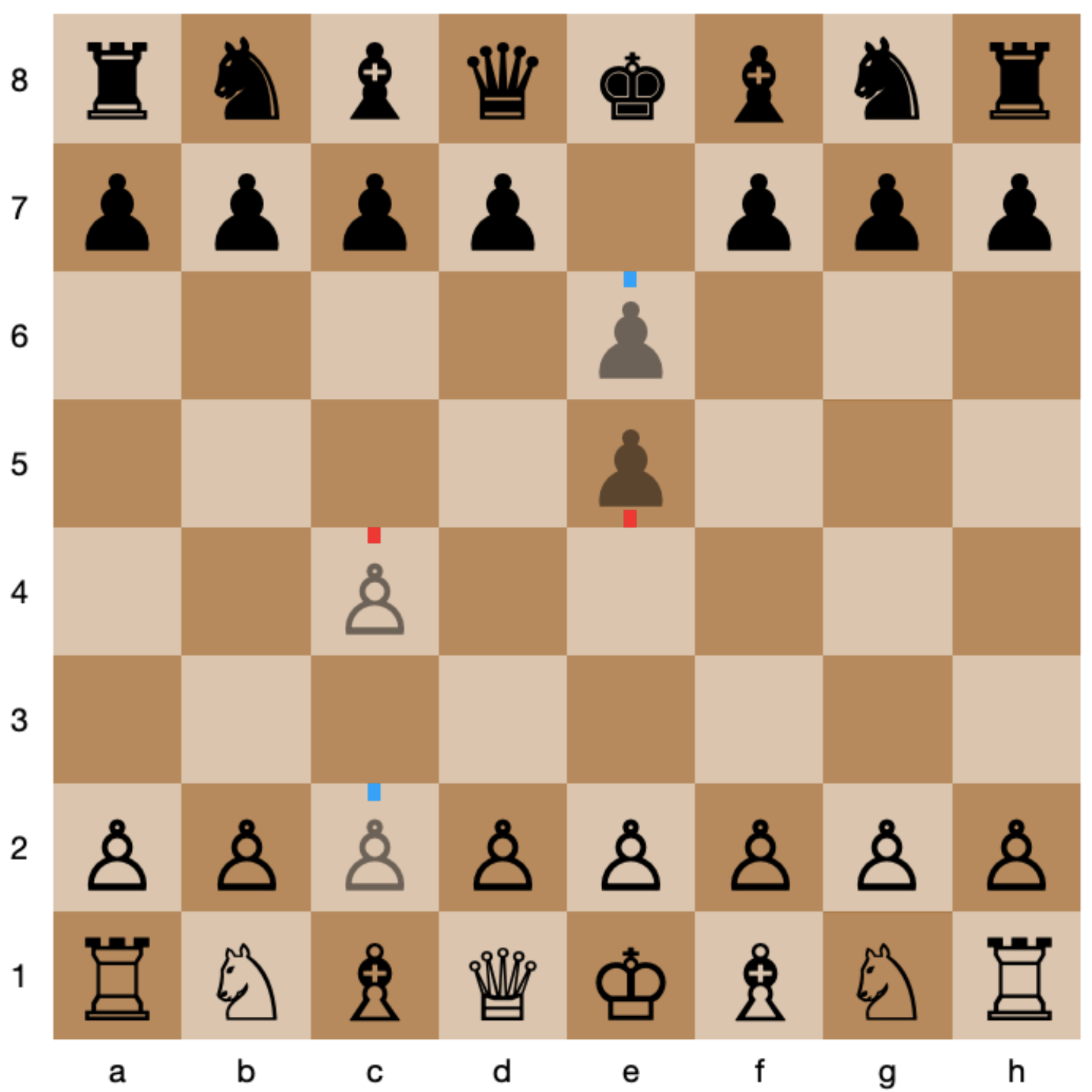}
\end{figure}
\item[7.4] A capture in which at least one indefinite pair instance is involved is called an 'attempted capture'.\\
\begin{enumerate}[1.1.1]
\item[7.4.1] \textit{[Clarification]} An indefinite piece may attempt to capture not only another indefinite piece but also a conventional piece, and vice versa.
\end{enumerate}
\begin{figure}[H]
\centering
\includegraphics[width=0.9\columnwidth]{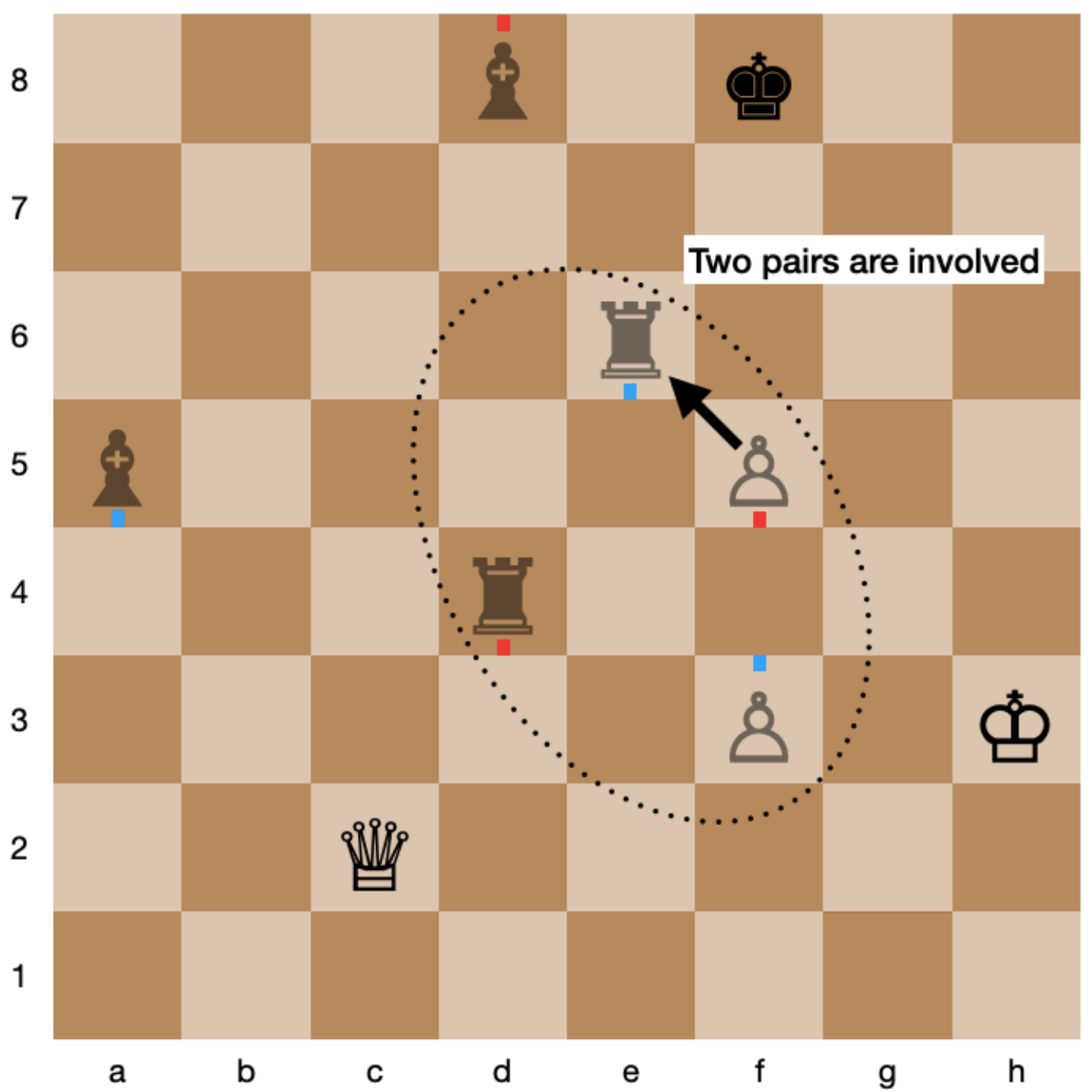}
\end{figure}
\item[7.5] Steps to execute the attempted capture (see also Article 8.4):\\
\begin{enumerate}[1.]
\item[1)] The player attempting the capture places his/her piece on the intended square of arrival.
\begin{figure}[H]
\centering
\includegraphics[width=0.9\columnwidth]{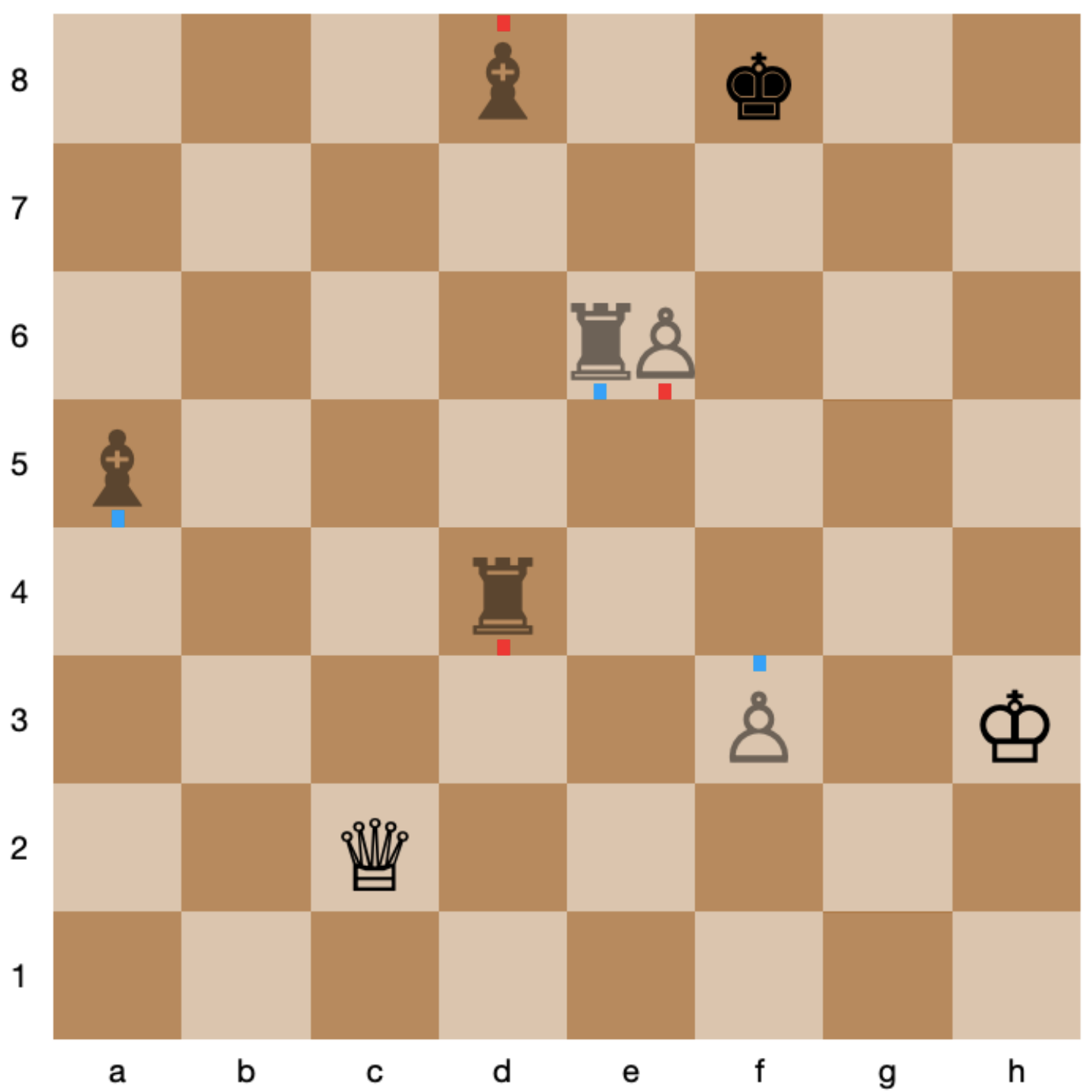}
\end{figure}
\item[2)] Each player who has an indefinite piece on the intended square of arrival rolls a six-sided dice:\\\\
- \textbf{Equal superposition with opponent-facing marks: }an even result 2, 4 or 6 (an odd result 1, 3 or 5) means that the player's indefinite pair instance 'collapses' to the square having the indefinite piece with the red mark (the blue mark) on it.\\\\
- \textbf{Equal superposition with player-facing marks: }same instructions as for the opponent-facing case, except that the words "red" and "blue" are swapped.\\\\
- \textbf{Unequal superposition: }a result 1, 2, 3 or 4 (a result 5 or 6) means that the player's indefinite pair instance 'collapses' to the square having the indefinite piece on it whose mark faces the opponent (the player).\\\\
- The two indefinite pieces are \textbf{replaced by the corresponding conventional piece} that must be placed on the square to which the pair has collapsed (as collapse means that the conventional piece has a definite position now).\\\\
- \textbf{Order: }if both players have to roll the dice, the one whose piece is attempted to be captured goes first.
\begin{figure}[H]
\centering
\includegraphics[width=0.9\columnwidth]{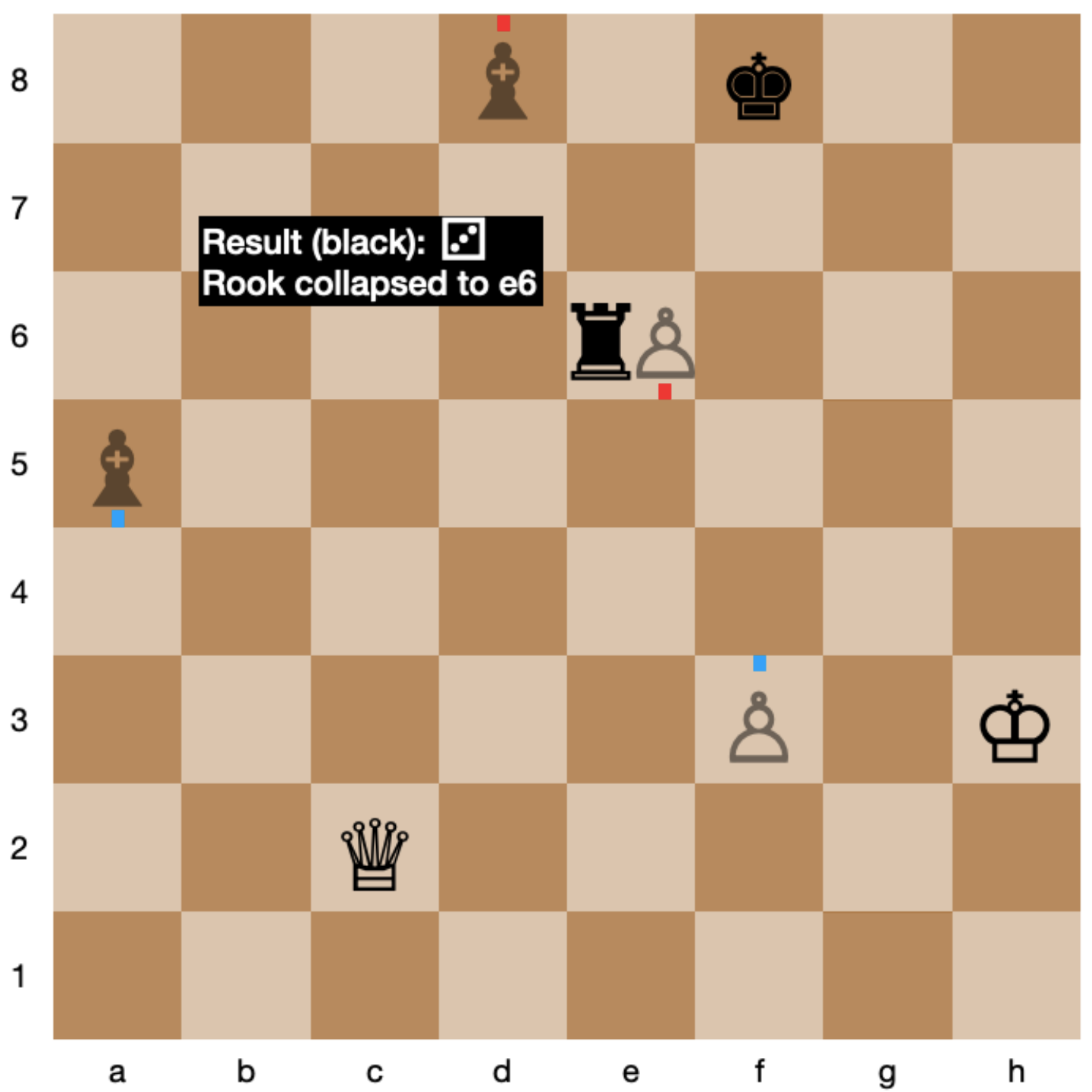}
\vspace*{-0.5cm}
\end{figure}
\begin{figure}[H]
\centering
\includegraphics[width=0.9\columnwidth]{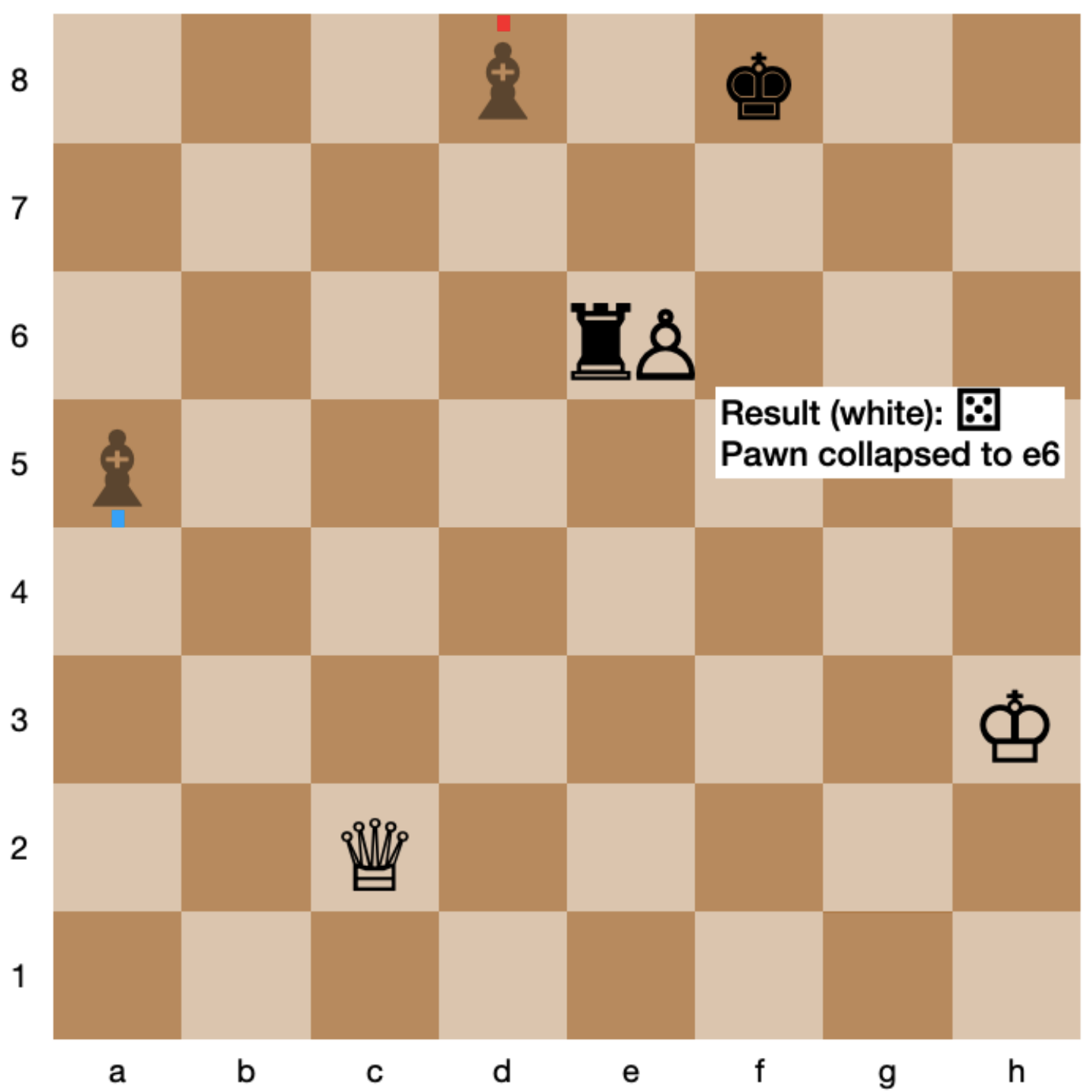}
\end{figure}
\item[3)] If, after step 2, two conventional pieces are on the same square, the capture is said to be 'successful': the piece owned by the player attempting the capture stays, while the other is \textbf{captured and removed from the chessboard}.\\
\end{enumerate}
\begin{figure}[H]
\centering
\includegraphics[width=0.9\columnwidth]{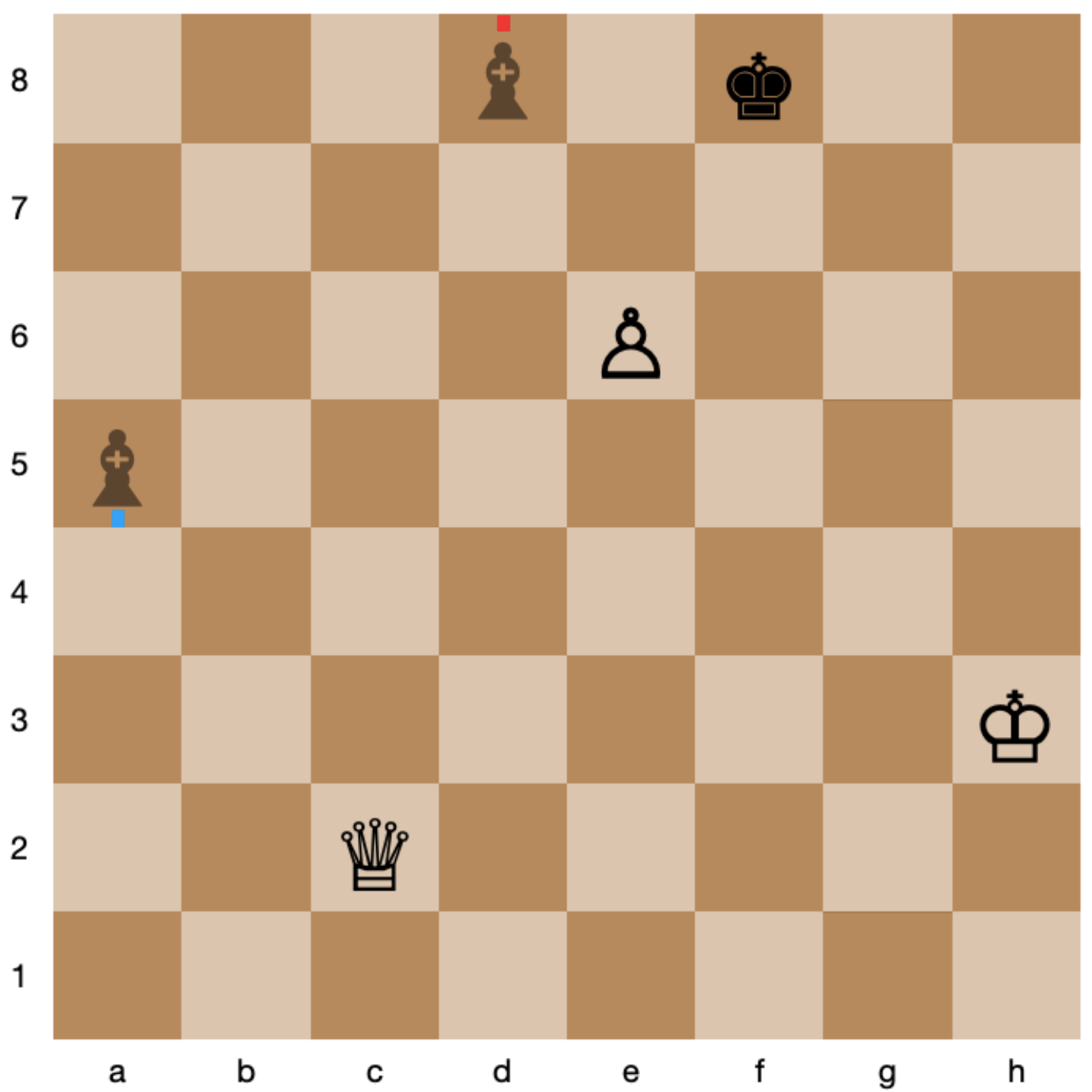}
\end{figure}
\begin{enumerate}[1.1.1]
\item[7.5.1] \textit{[Explanation]} Even the slightest interaction with the environment can cause a superposition state to collapse. That's why, an indefinite piece is meant to be surrounded by a protection layer (sealed box). However, when it hits, or is hit by, another piece, indefinite or not, the isolation gets broken and the superposition state collapses immediately.\\
\end{enumerate}
\item[7.6] It is not allowed to make a move that could potentially expose the king, indefinite or not, of the same colour to check or leave that king in check (by one or more of the opponent's conventional or indefinite pieces).
\begin{figure}[H]
\centering
\includegraphics[width=0.9\columnwidth]{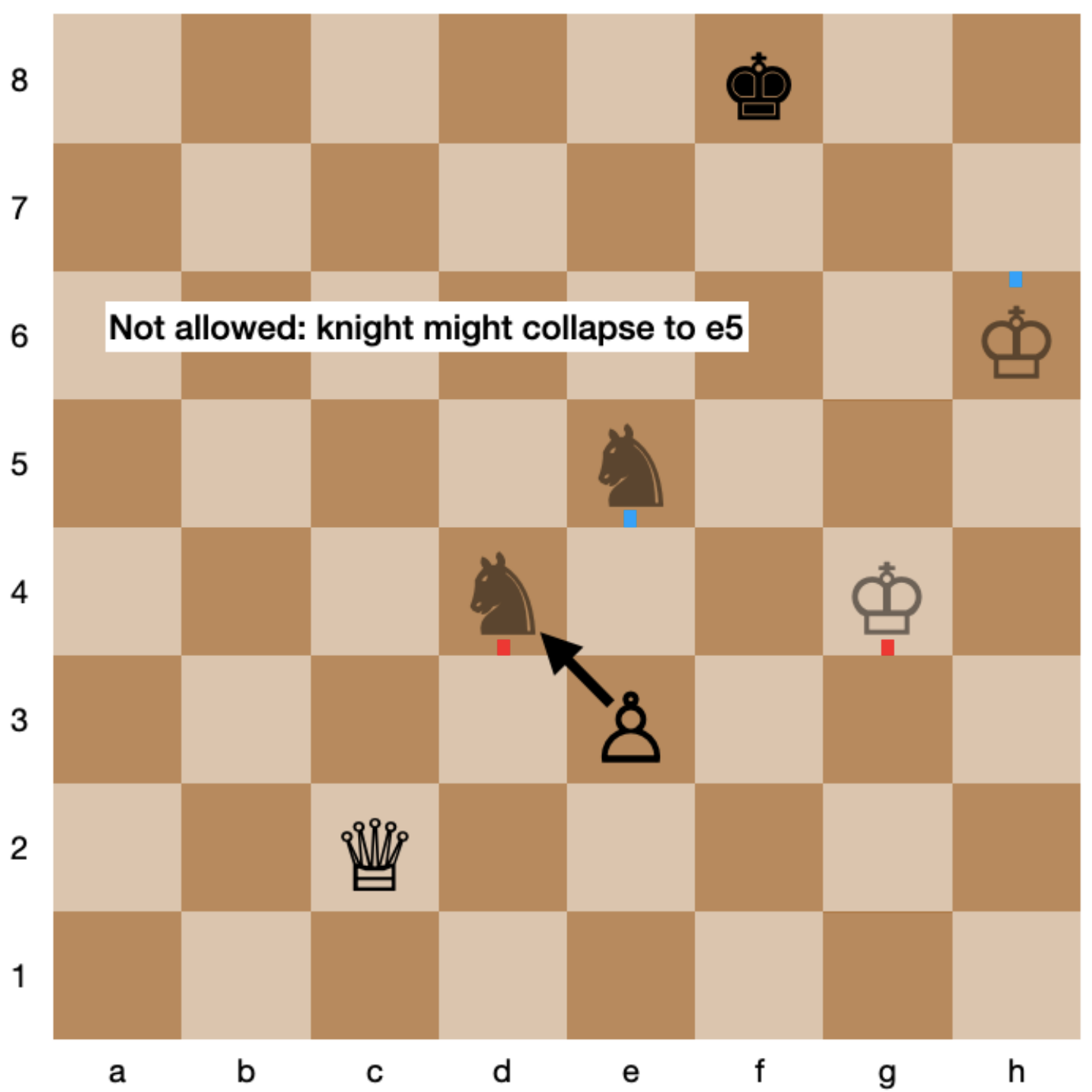}
\end{figure}
\item[7.7] If an indefinite piece is 'one move away' from its paired piece (i.e. it could move there if the square on which its paired piece stands was unoccupied, not considering whether such a move would expose the king of the same colour to check or leave that king in check), then the player who owns the pair may make a move that merely inverts the facing of the marks on both pieces (by rotating each piece by $180^\circ$), provided that it makes a real difference to the position.\\
\begin{figure}[H]
\centering
\includegraphics[width=0.9\columnwidth]{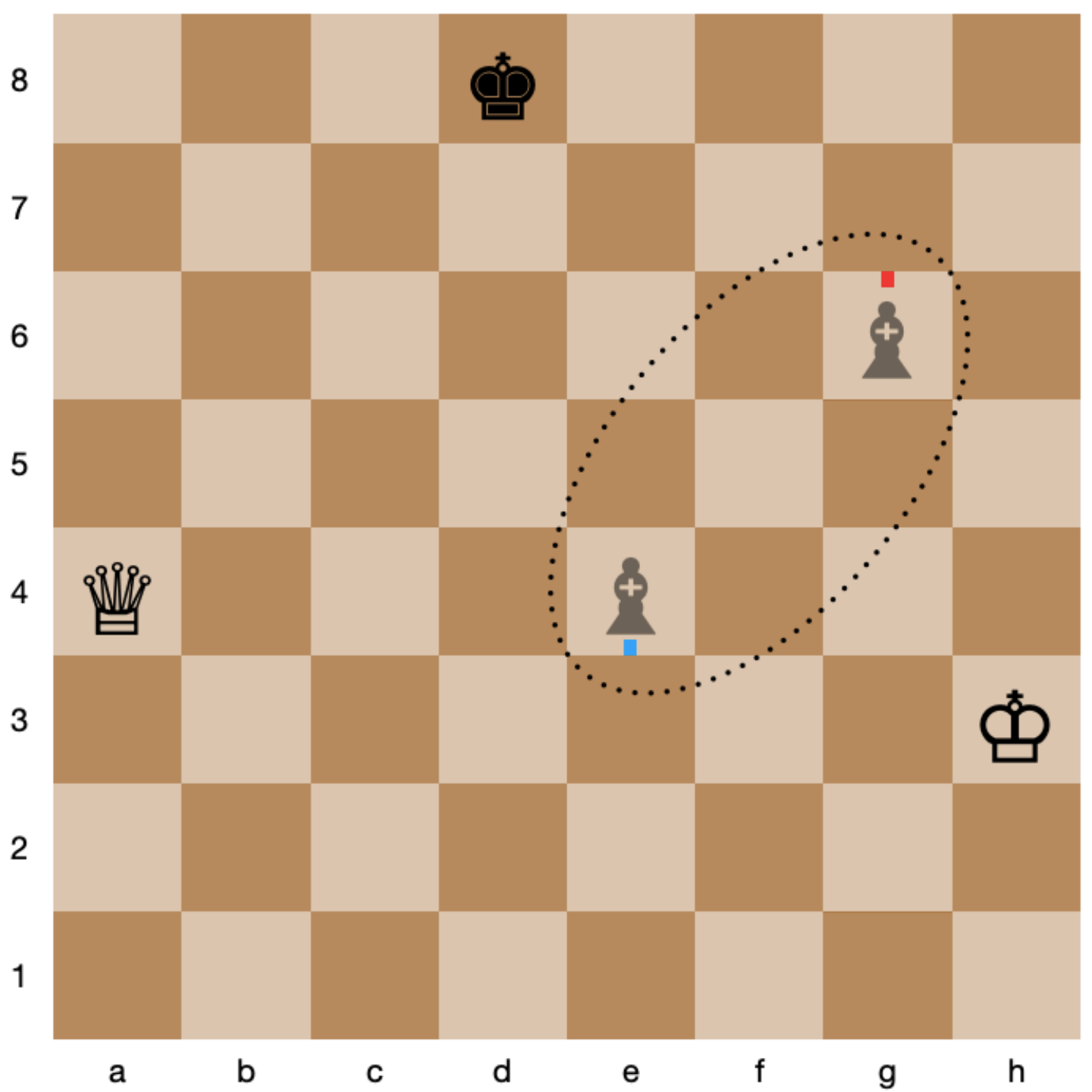}
\vspace*{-0.5cm}
\end{figure}
\begin{figure}[H]
\centering
\includegraphics[width=0.9\columnwidth]{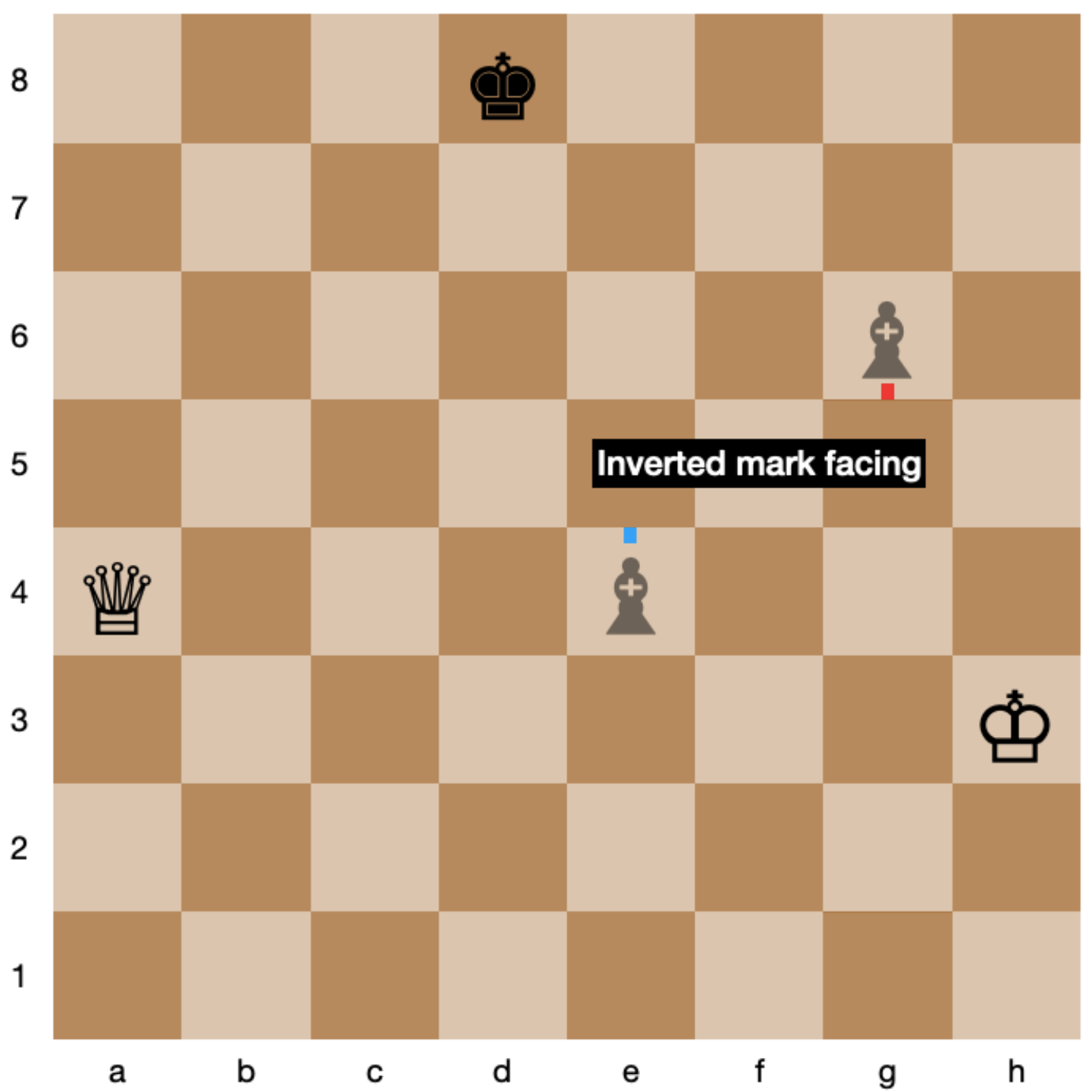}
\end{figure}
\item[7.8] If an indefinite pawn would occupy the furthest rank after a move, the indefinite pair instance that it belongs to must be exchanged as part of the same move for a new indefinite queen, rook, bishop or knight pair instance of the same colour and of the same facing of the marks.
\begin{figure}[H]
\centering
\includegraphics[width=0.9\columnwidth]{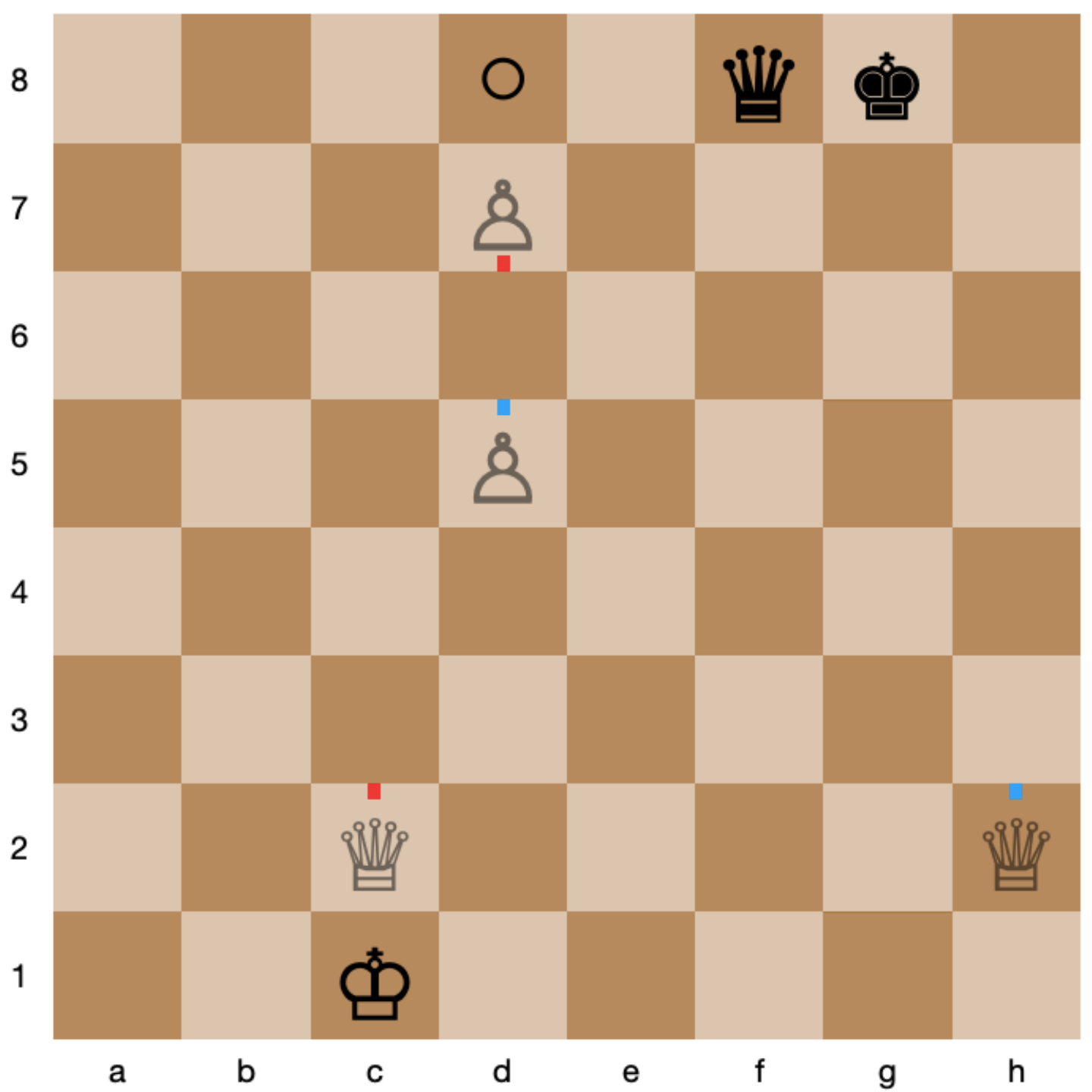}
\vspace*{-0.5cm}
\end{figure}
\begin{figure}[H]
\centering
\includegraphics[width=0.9\columnwidth]{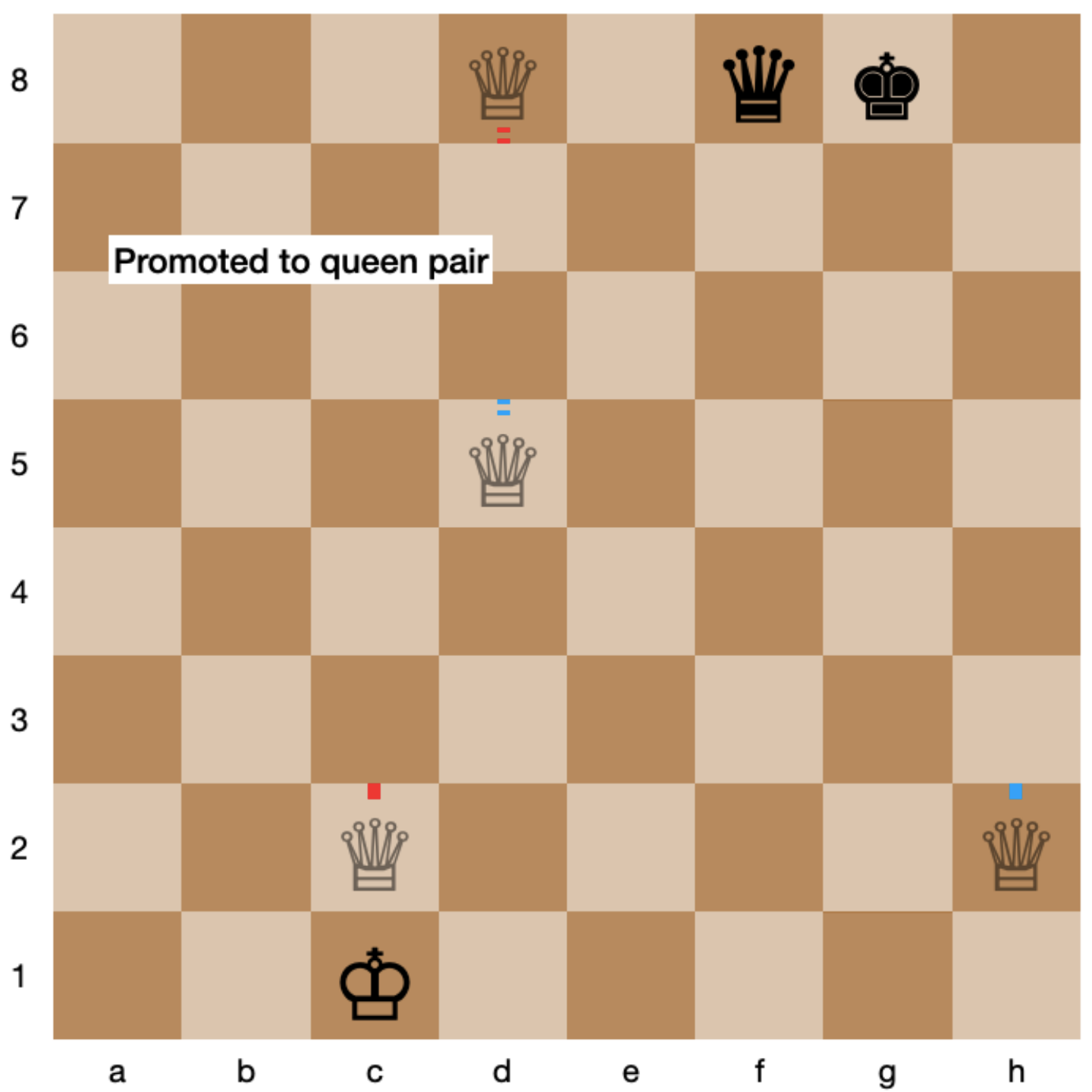}
\end{figure}
\item[7.9] \textit{[Clarification]} In accordance with Articles 3.7.3.1 and 3.7.3.2, the 'en passant' capture of an indefinite pawn which has just advanced two squares in a superposition first move must be made as though it had moved only one square forward (while its pair had stayed in the starting position).\\
\end{enumerate}

\section{More on the indefinite checkmate}\label{spnotes}

In Niel's Chess, it is possible to checkmate the opponent's king by placing it under 'indefinite attack'. A couple of examples are shown below, followed by a rationale behind such situations in general.

\subsection*{Indefinite checkmate examples}

The positions here may look unusual for players of conventional chess.

\begin{figure}[H]
\centering
\includegraphics[width=0.9\columnwidth]{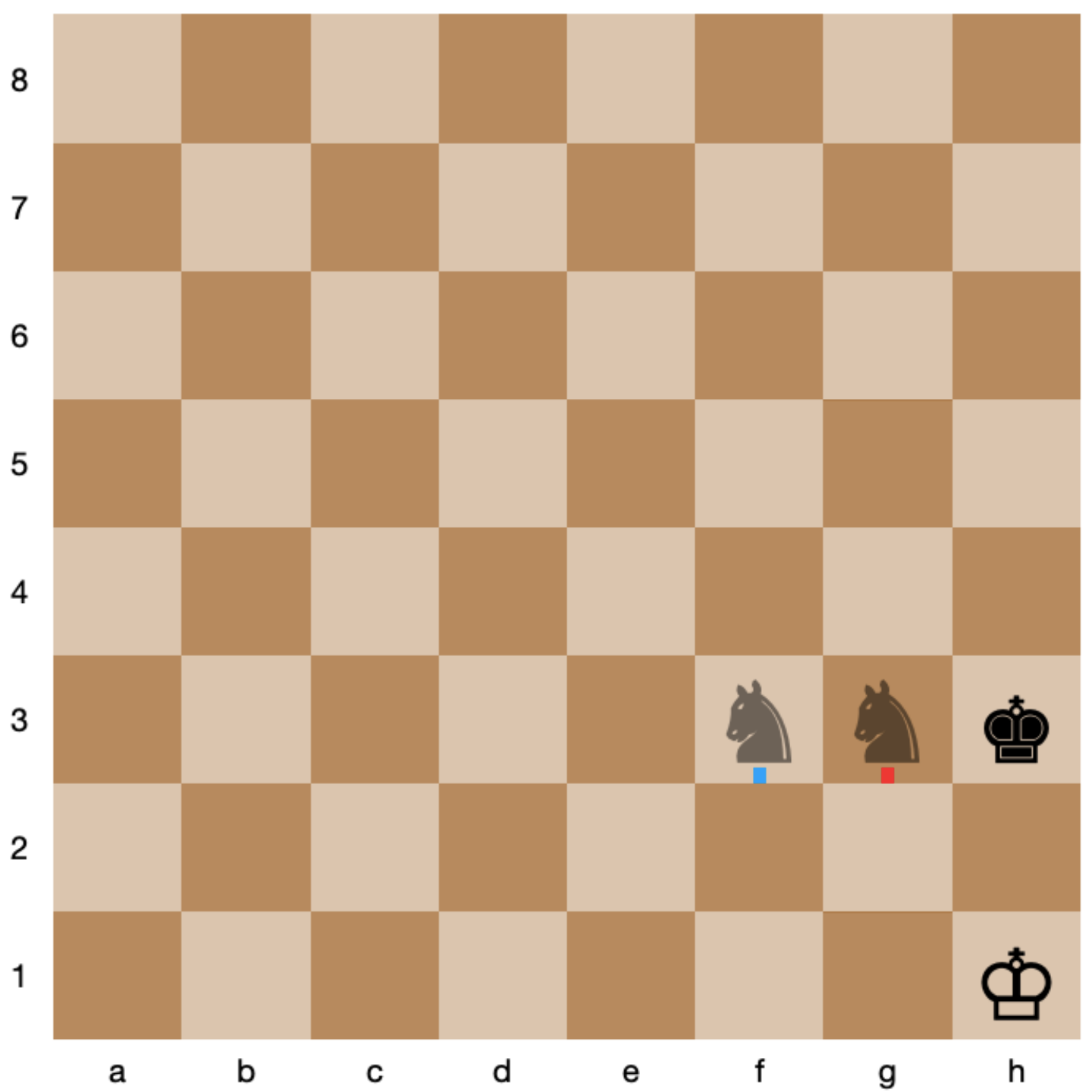}
\vspace*{-0.5cm}
\end{figure}
\begin{figure}[H]
\centering
\includegraphics[width=0.9\columnwidth]{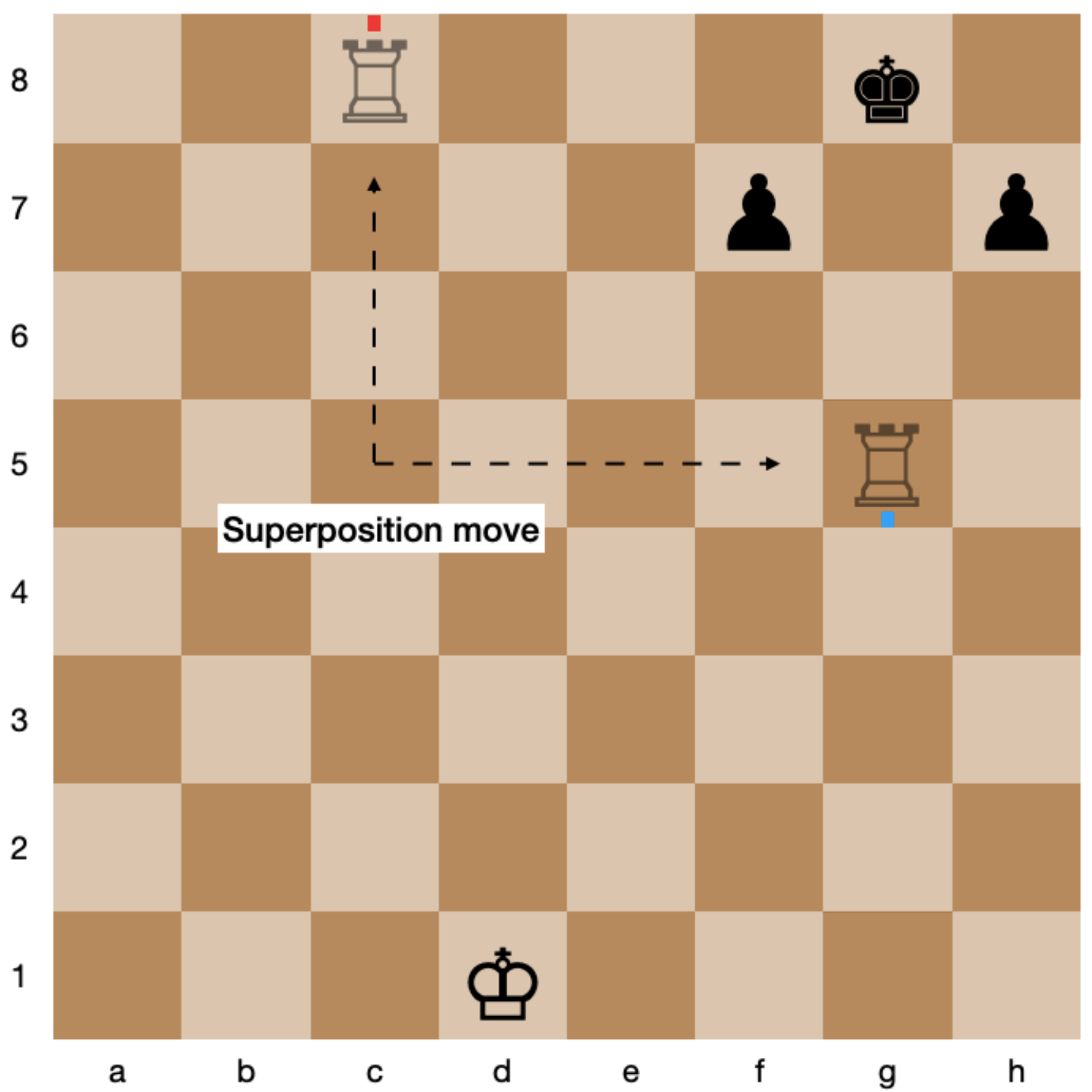}
\vspace*{-0.50cm}
\end{figure}
\begin{figure}[H]
\centering
\includegraphics[width=0.9\columnwidth]{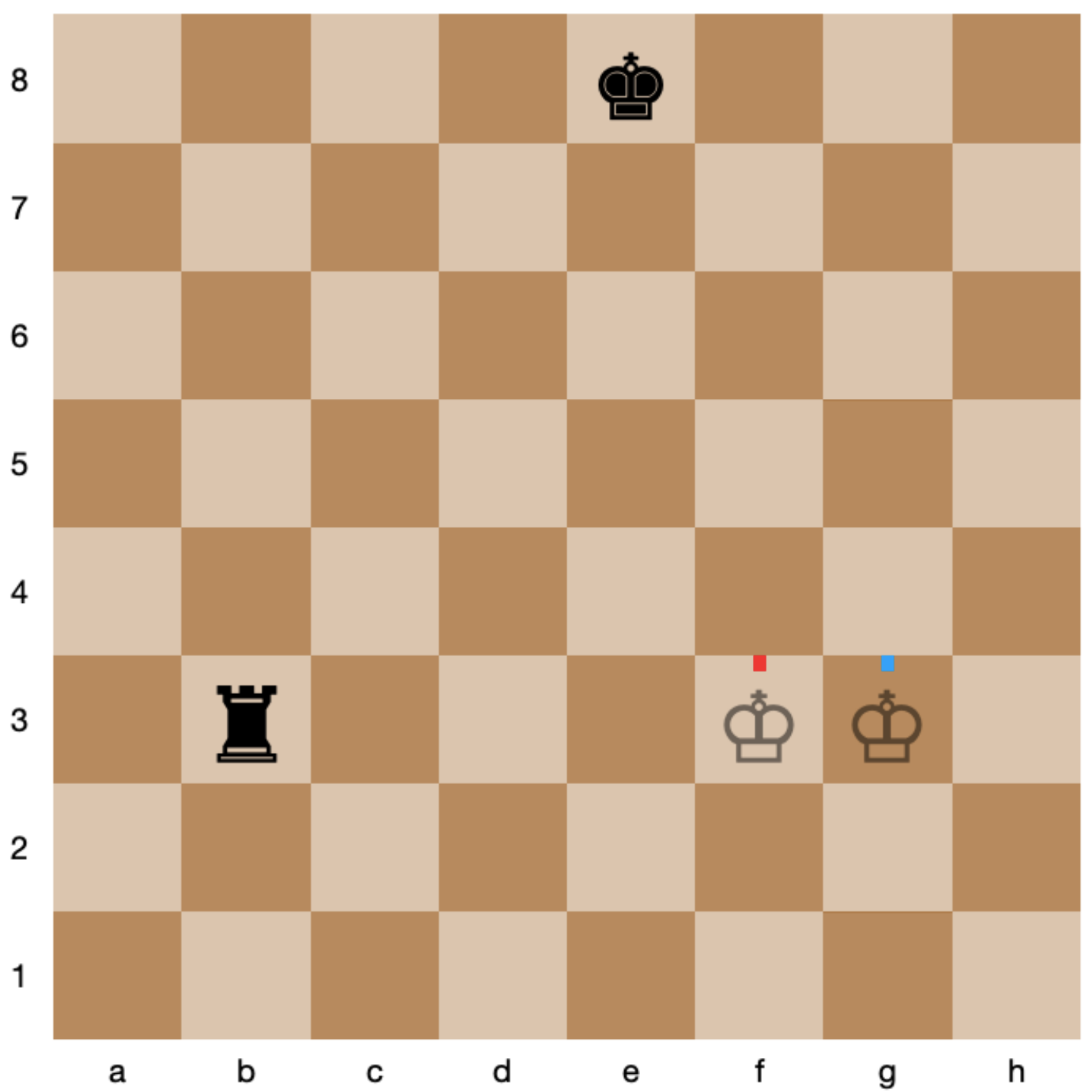}
\vspace*{-0.51cm}
\end{figure}
\begin{figure}[H]
\centering
\includegraphics[width=0.9\columnwidth]{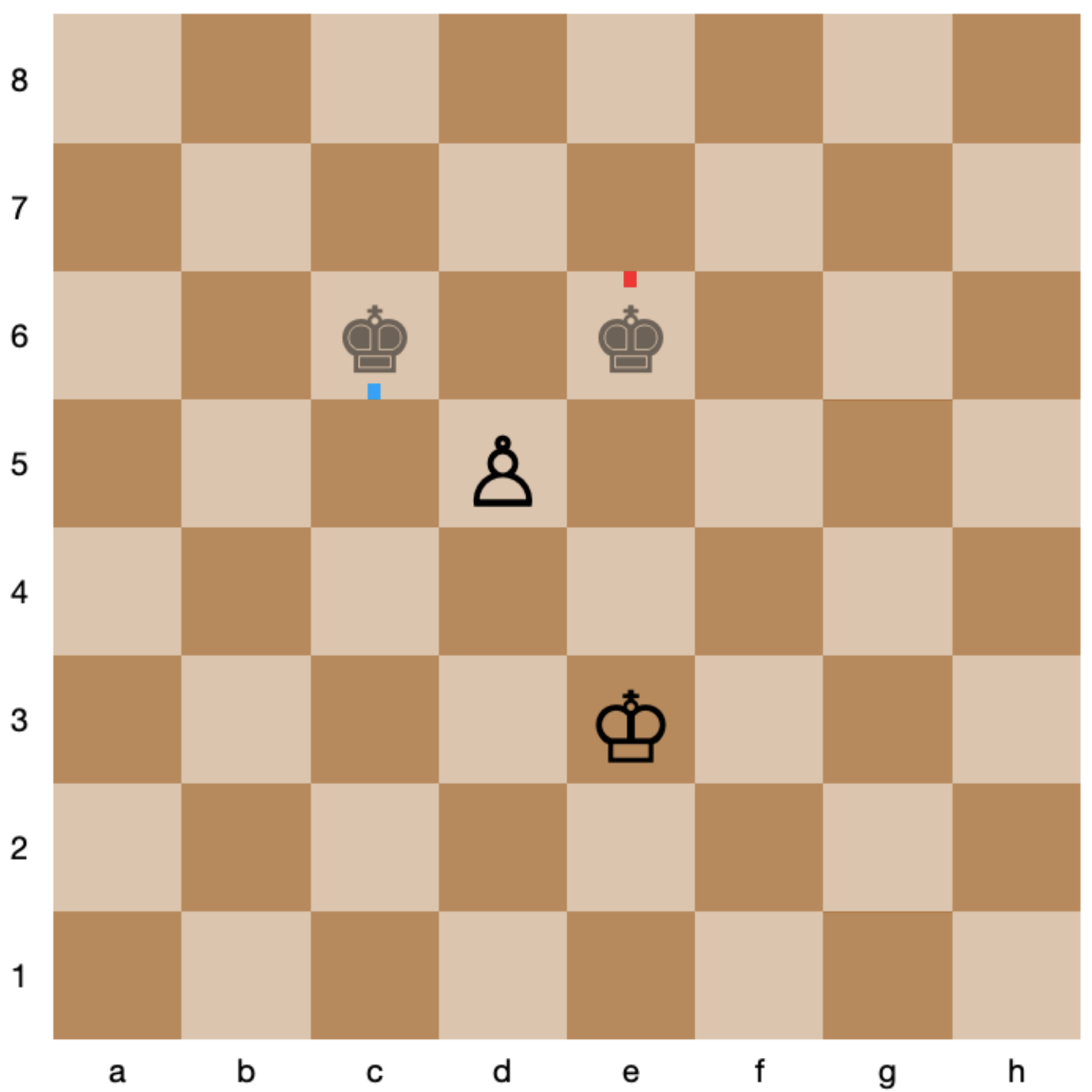}
\end{figure}

\subsection*{Why is it a checkmate?}

One may argue that in the examples above, the king wouldn't be captured with certainty in a hypothetical next move, only with some non-zero probability, so the king should be given a chance to escape, instead of ending the game.

Calling such positions a checkmate is inspired by quantum computing. Due to the probabilistic nature of quantum physics (plus the noise in the quantum hardware), it is typical that a quantum program outputs the correct answer to a problem not with certainty but only with some non-negligible probability. And that, provided that the solution can be easily verified, already suffices to call the problem 'solved'.

For example, if there is a $50\%$ chance of getting the correct answer, then by executing the program as few as 10 times, there will be more than a $99.9\%$ chance of seeing the correct answer at least once. Additional runs would get the user arbitrarily close to certainty.

So, in Niel's Chess, the goal is to place the opponent’s king ‘under risk’ in a way that no move can guarantee to fully eliminate it.

\subsection*{On the role of luck}

In principle, it may happen that most or even all of the $64$ squares are occupied by indefinite pieces, resulting in chaotic positions with plenty of attempted captures whose outcomes are probabilistic.

It's easy to lose clarity and control in such situations and conclude that winning or losing is just a matter of luck. However, it's also possible that the player just hasn't yet developed an intuition for useful tactics and patterns in a game with quantum effects.

\section{Basic rules of play - Part 2}\label{basic2}

In case of conflict, Article 8 has priority over Article 7.

\subsection*{Article 8: Entanglement rules}

\begin{enumerate}[1.1]
\item[8.1] A conventional piece may join the superposition in which an opponent's indefinite piece participates. This is called an 'entanglement move'. The conventional piece has to be replaced by the two pieces of a corresponding indefinite pair instance, one placed on the square of departure, and the other on an unoccupied square of arrival (reachable as per Articles 3.1 to 3.8). At least one indefinite piece of the pair instance must attack the opponent's indefinite piece.\footnote{In the figure below, the dotted ellipse indicates that the knight on f6 attacks the rook on e4. Also, an indefinite knight on the square of arrival d5 would attack the rook on f4. Either of these two reasons is sufficient to allow an entanglement move involving the rook pair, as stated in Article 8.1.} (The point of entanglement is that the involved pairs become connected, in the sense that an attempted capture will collapse them in tandem, in a correlated fashion, see Article 8.4.)
\begin{figure}[H]
\centering
\includegraphics[width=0.9\columnwidth]{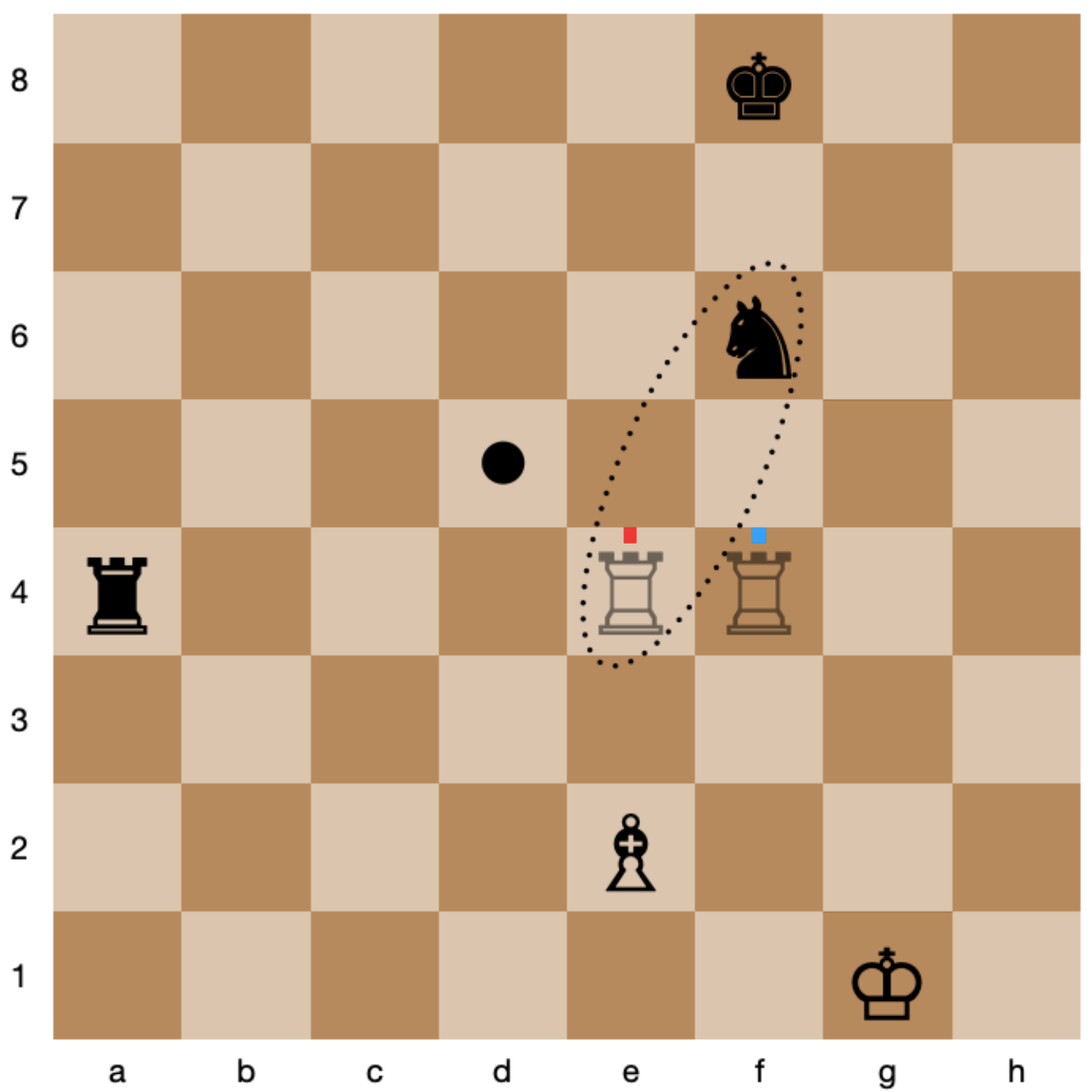}
\end{figure}
\item[8.2] The mark on each indefinite piece which hasn't yet been involved in an entanglement move must lie on the vertical mid-line of its square (just like in all figures so far).\\
\item[8.3] Steps to execute an entanglement move:\\
\begin{enumerate}[1.]
\item[1)] The player making the move directs the marks of his/her newly placed indefinite pair instance such that they lie on the vertical mid-line of the respective squares, indicating \textbf{the same type of superposition} (equal or unequal) as those of the opponent's pair involved.
\begin{figure}[H]
\centering
\includegraphics[width=0.9\columnwidth]{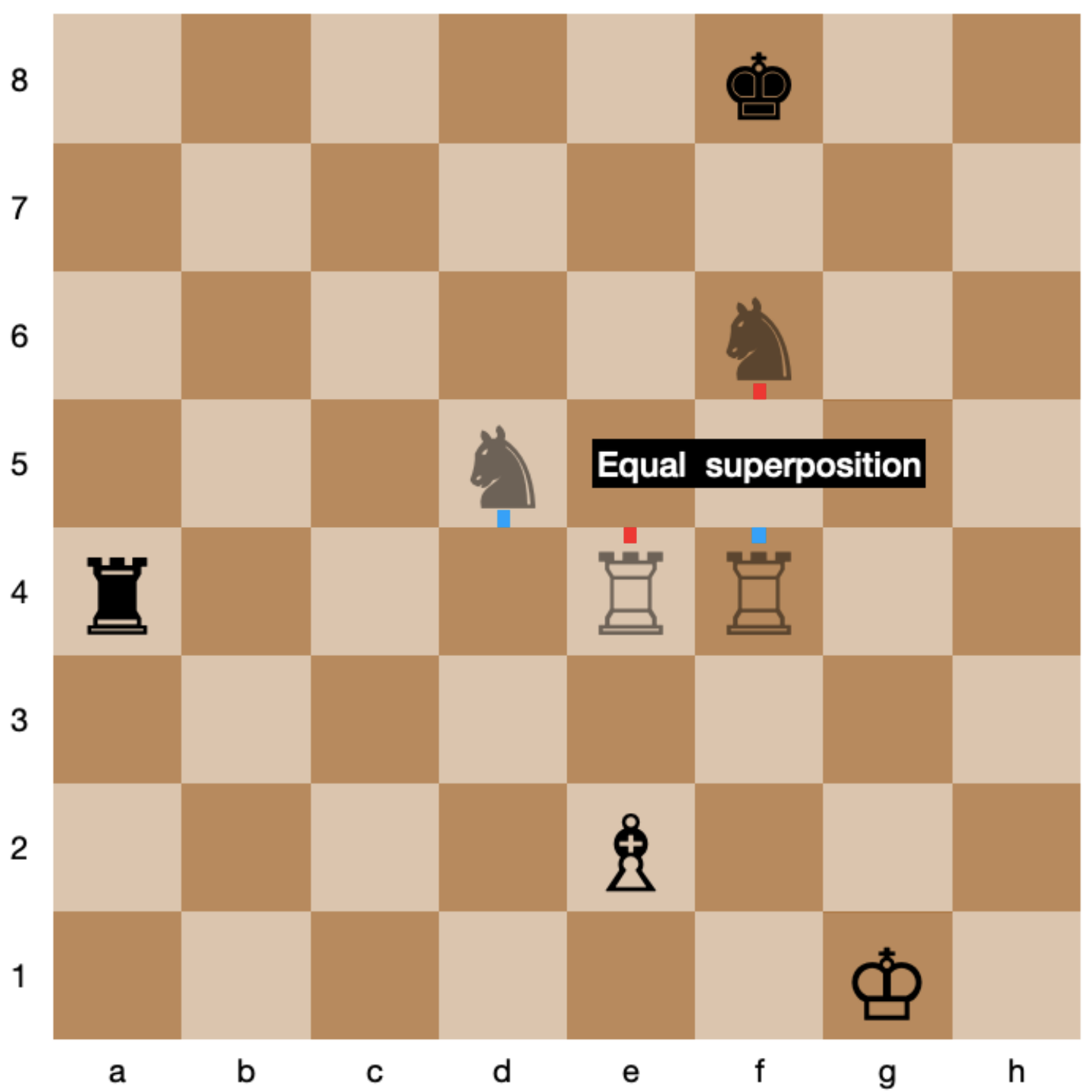}
\end{figure}
\item[2)] If the opponent's marks lie on the vertical mid-line, the opponent must \textbf{rotate both pieces by the same angle}, either $+45^\circ$ or $-45^\circ$, to make the marks lie on the same diagonal of their respective squares. A diagonal cannot be chosen if there are already other indefinite pieces whose marks lie on the same diagonal. If neither of the diagonals can be chosen, the entanglement move \textbf{is not allowed}.
\begin{figure}[H]
\centering
\includegraphics[width=0.9\columnwidth]{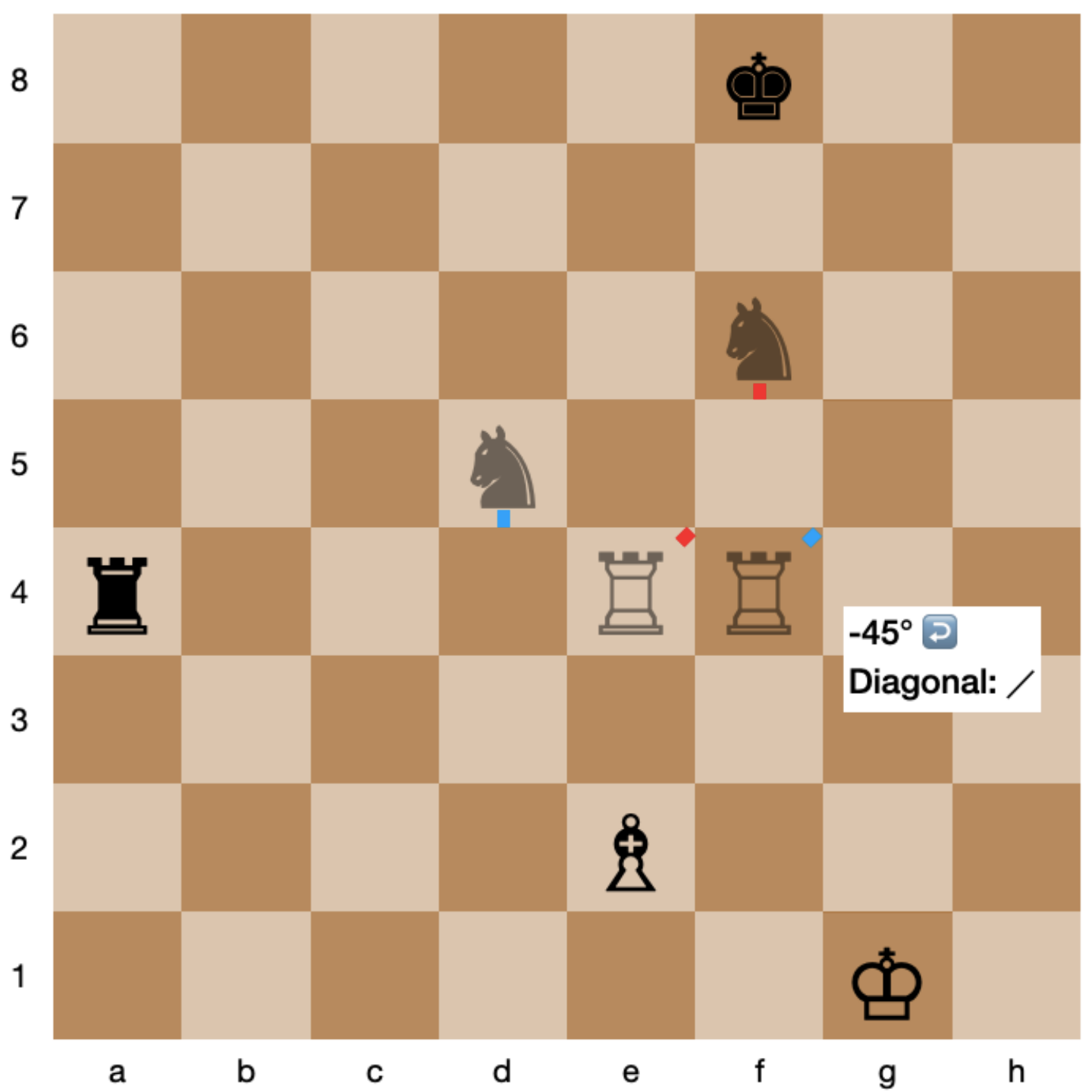}
\end{figure}
\item[3)] The player making the move rotates his/her pieces to align their marks with those of the opponent's.
\begin{figure}[H]
\centering
\includegraphics[width=0.9\columnwidth]{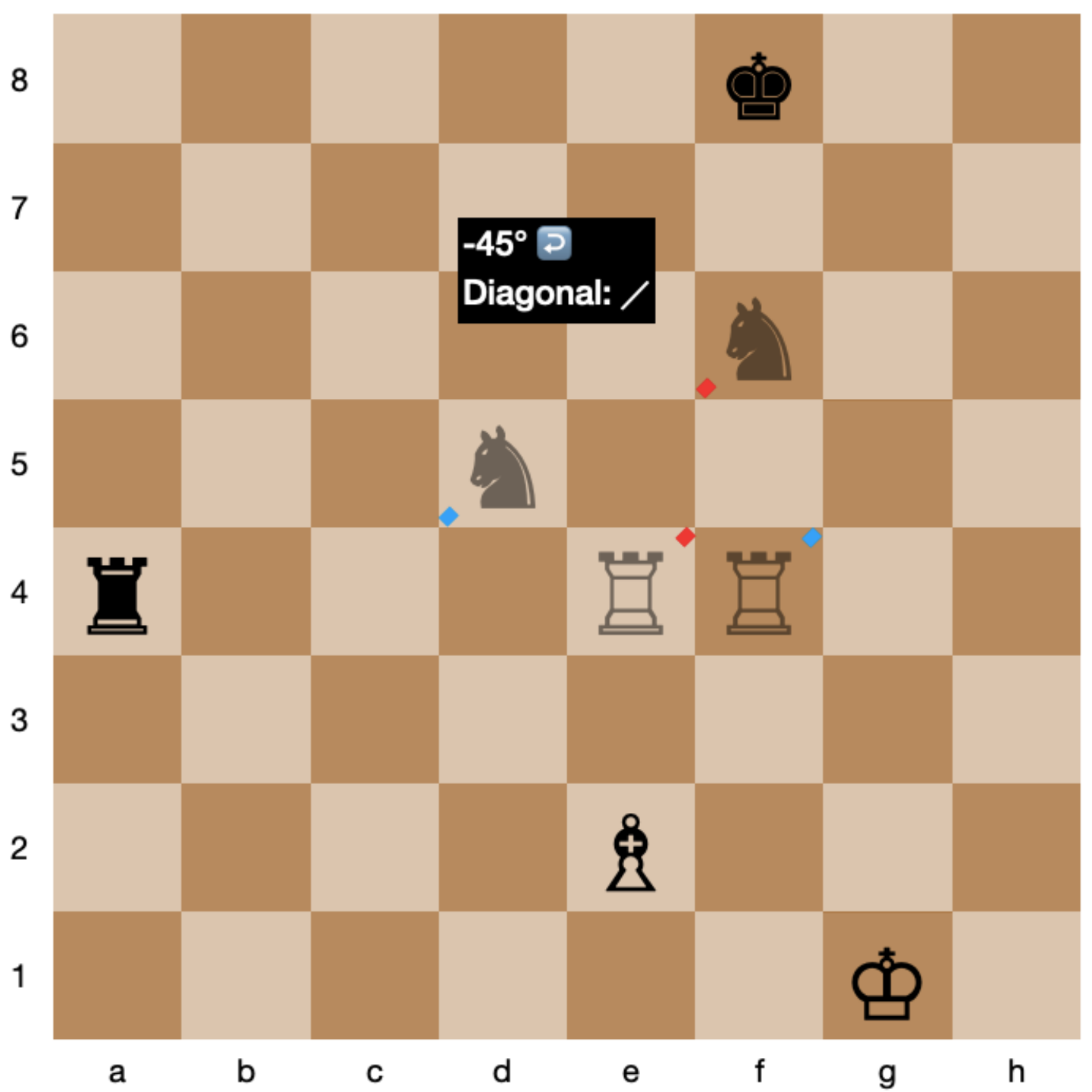}
\end{figure}
\end{enumerate}
\begin{enumerate}[1.1.1]
\item[8.3.1] \textit{[Clarification]} In accordance with Article 7.2, the rotations don't change the type of any superposition, as every mark will face the same upper or lower side of the square as before.\\
\item[8.3.2] \textit{[Explanation]} To justify the limitation imposed by step 2, it is noted that in quantum information, entanglement is a \textbf{resource}, which can be scarce. That's why, players are allowed only limited access to it.\\
\item[8.3.3] \textit{[Clarification]} Additional conventional pieces may join as well (with step 2 omitted).
\begin{figure}[H]
\centering
\includegraphics[width=0.9\columnwidth]{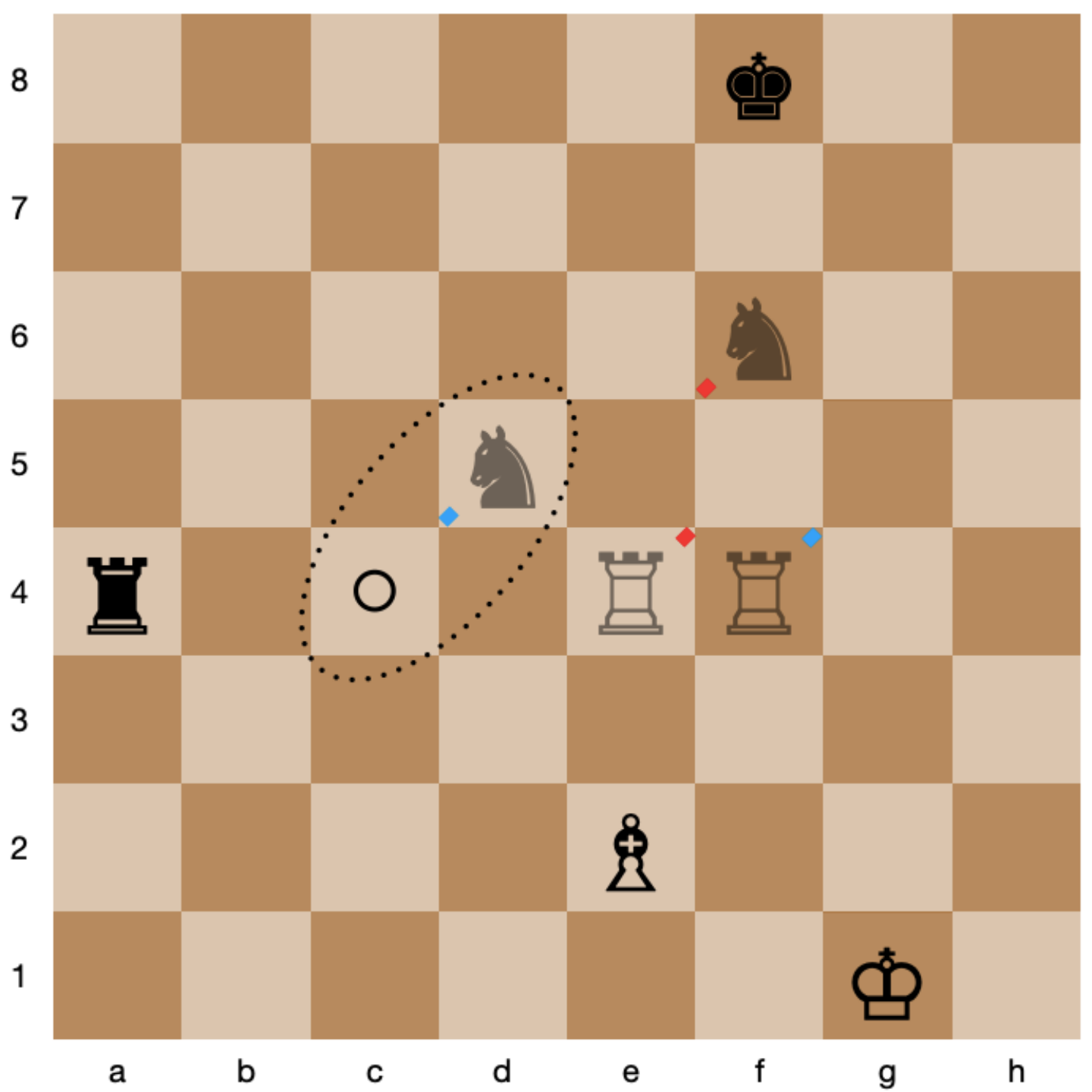}
\vspace*{-0.5cm}
\end{figure}
\begin{figure}[H]
\centering
\includegraphics[width=0.9\columnwidth]{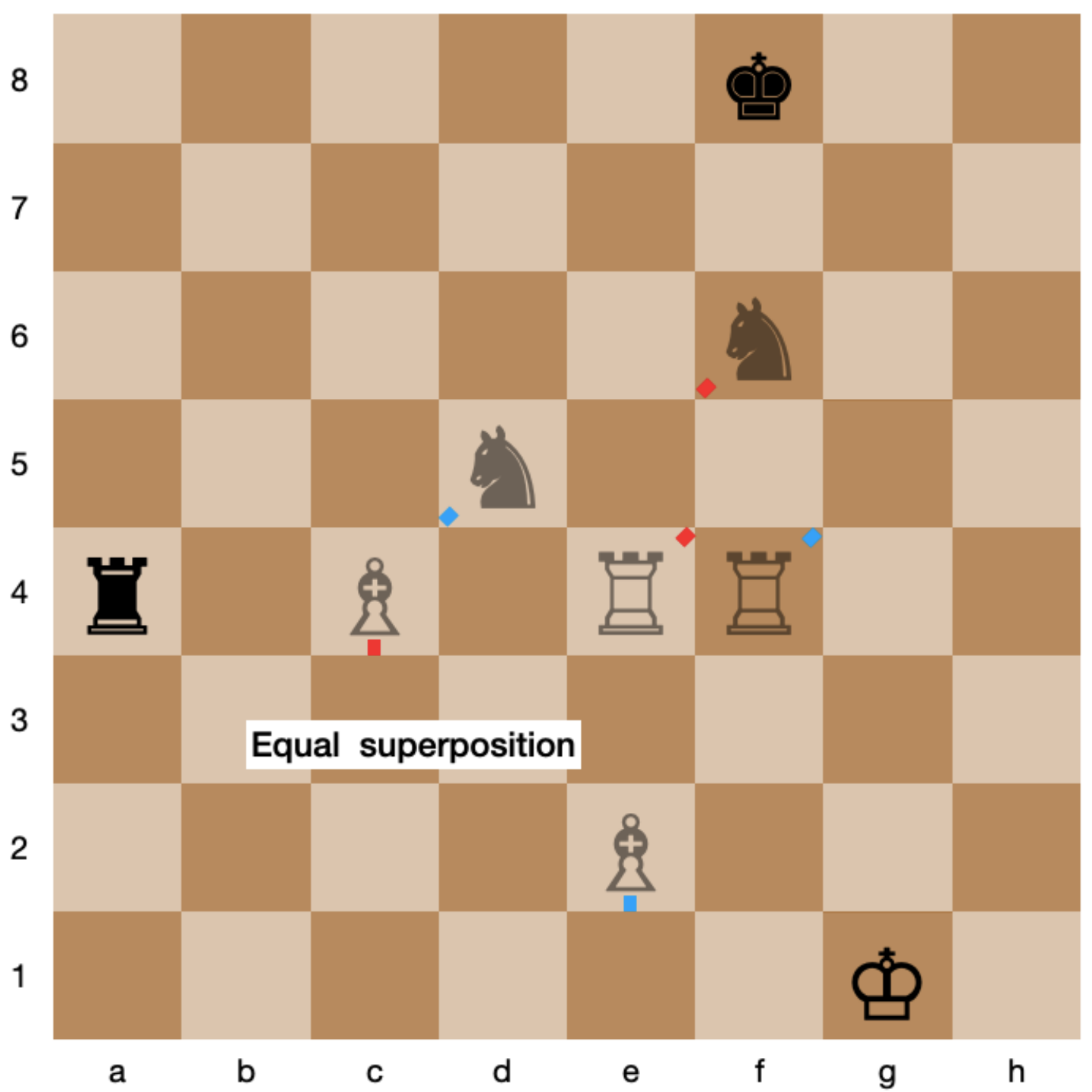}
\vspace*{-0.5cm}
\end{figure}
\begin{figure}[H]
\centering
\includegraphics[width=0.9\columnwidth]{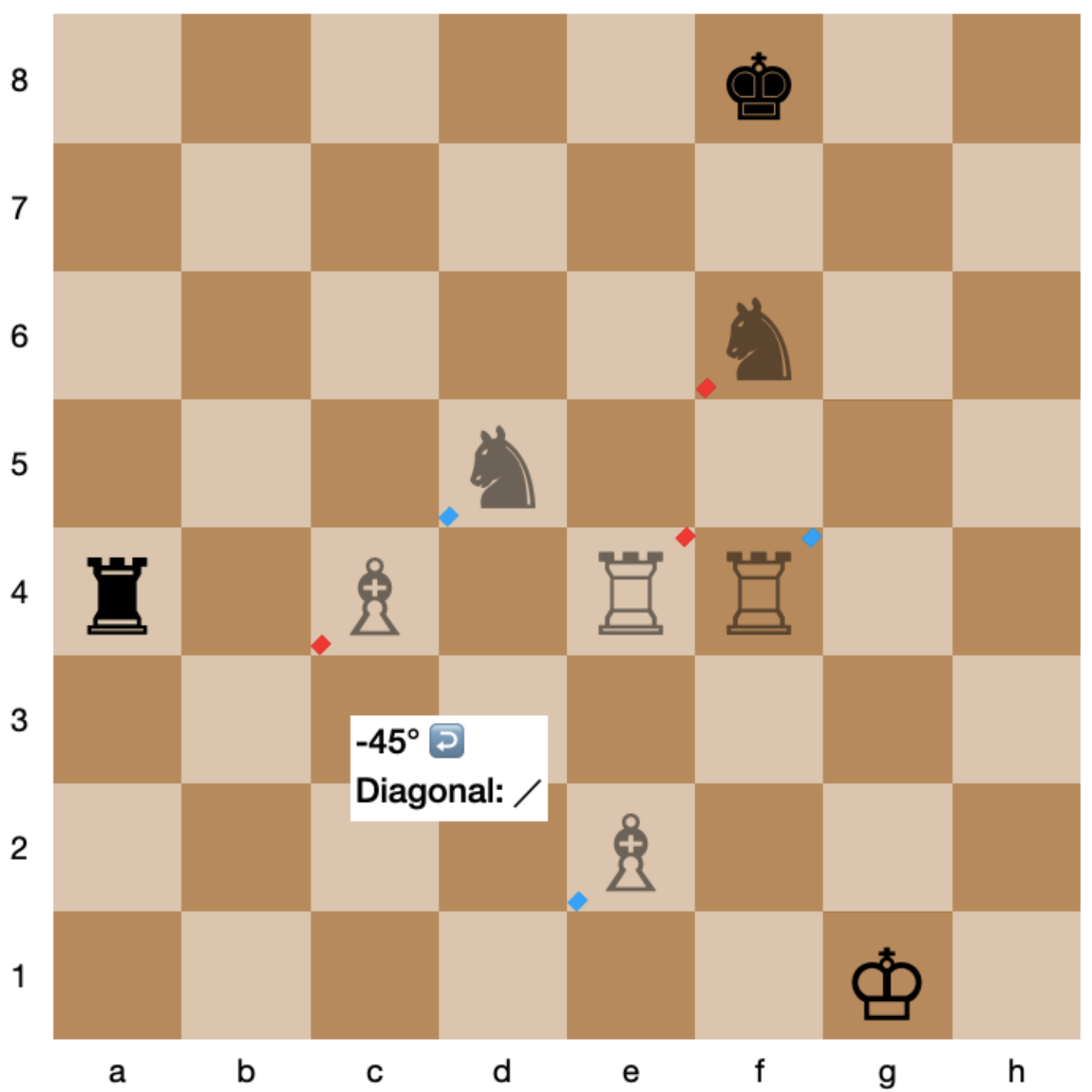}
\end{figure}
\item[8.3.4] \textit{[Clarification]} The indefinite pair instances whose marks lie on the same diagonal are said to be 'entangled' (as they represent the corresponding group of conventional pieces in an entangled state of correlated behaviour, see Article 8.4). Since a square has two diagonals, there can be at most two entangled groups of pairs on the chessboard. (Using a more complex scheme of directing the marks would enable the maximum of $16$ groups of entangled pairs. However, it would likely reduce the playability of the game.)
\begin{figure}[H]
\centering
\includegraphics[width=0.9\columnwidth]{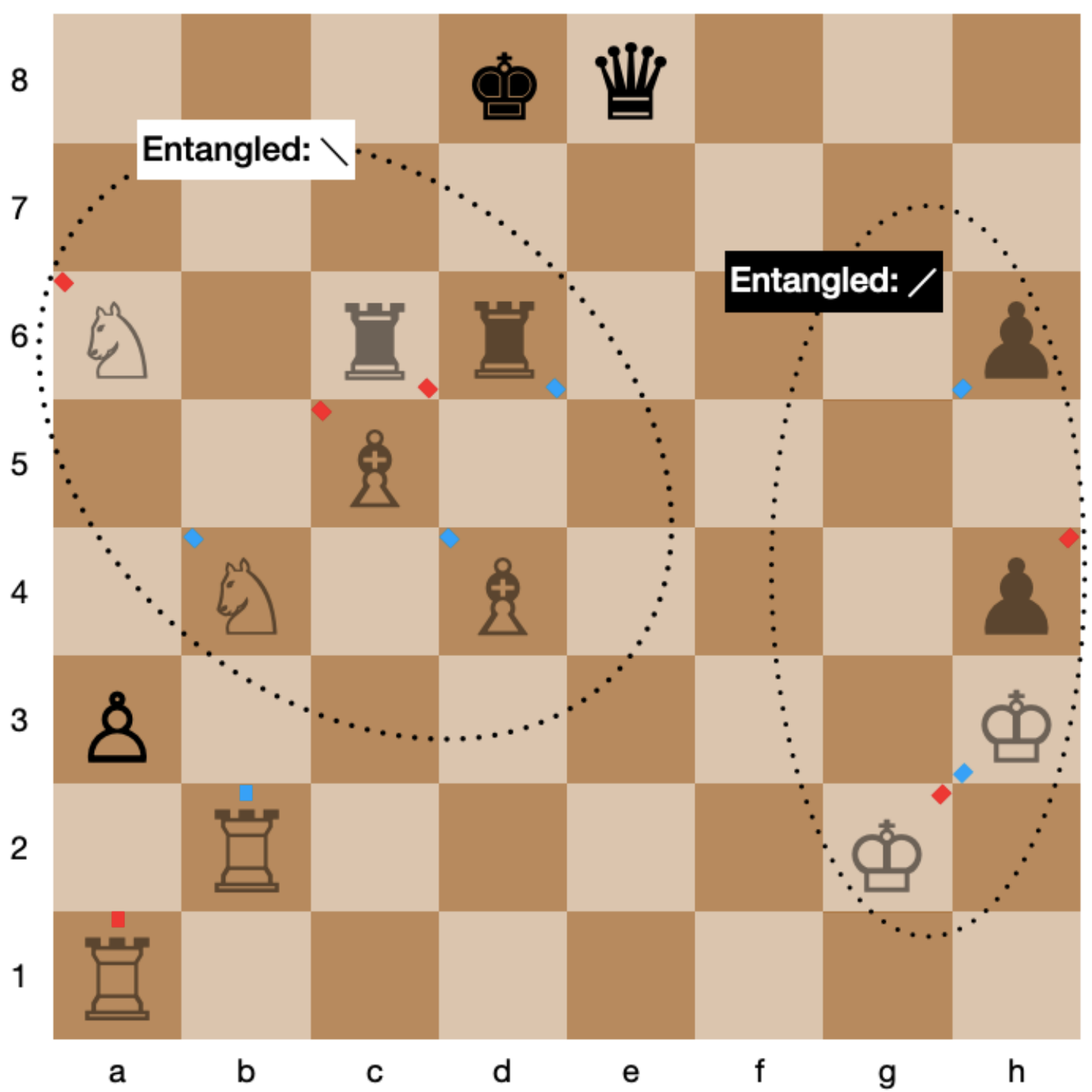}
\end{figure}
\end{enumerate}
\item[8.4] In step 2 of Article 7.5, right after collapsing an indefinite pair following a roll of the dice, all indefinite pair instances entangled with that pair must be collapsed as well, using the same result of the roll.\\
\begin{enumerate}[1.1.1]
\item[8.4.1] \textit{[Explanation]} Thus, rolling the dice results (randomly) in one of two possible collapse outcomes for the whole group of entangled pairs. This is because physically, \textbf{entanglement is a superposition} of those two possible holistic outcomes.
\end{enumerate}
\begin{figure}[H]
\centering
\includegraphics[width=0.9\columnwidth]{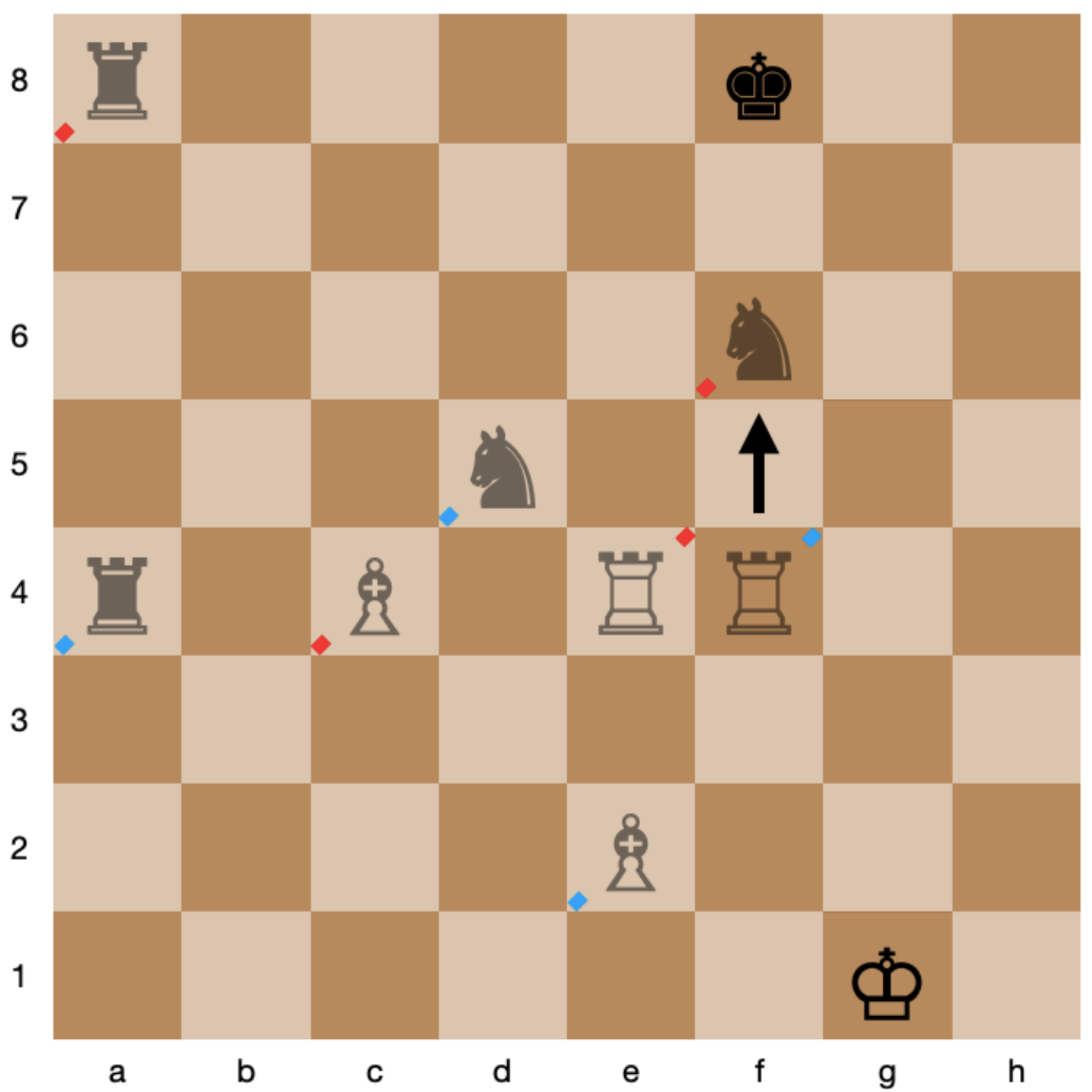}
\vspace*{-0.5cm}
\end{figure}
\begin{figure}[H]
\centering
\includegraphics[width=0.9\columnwidth]{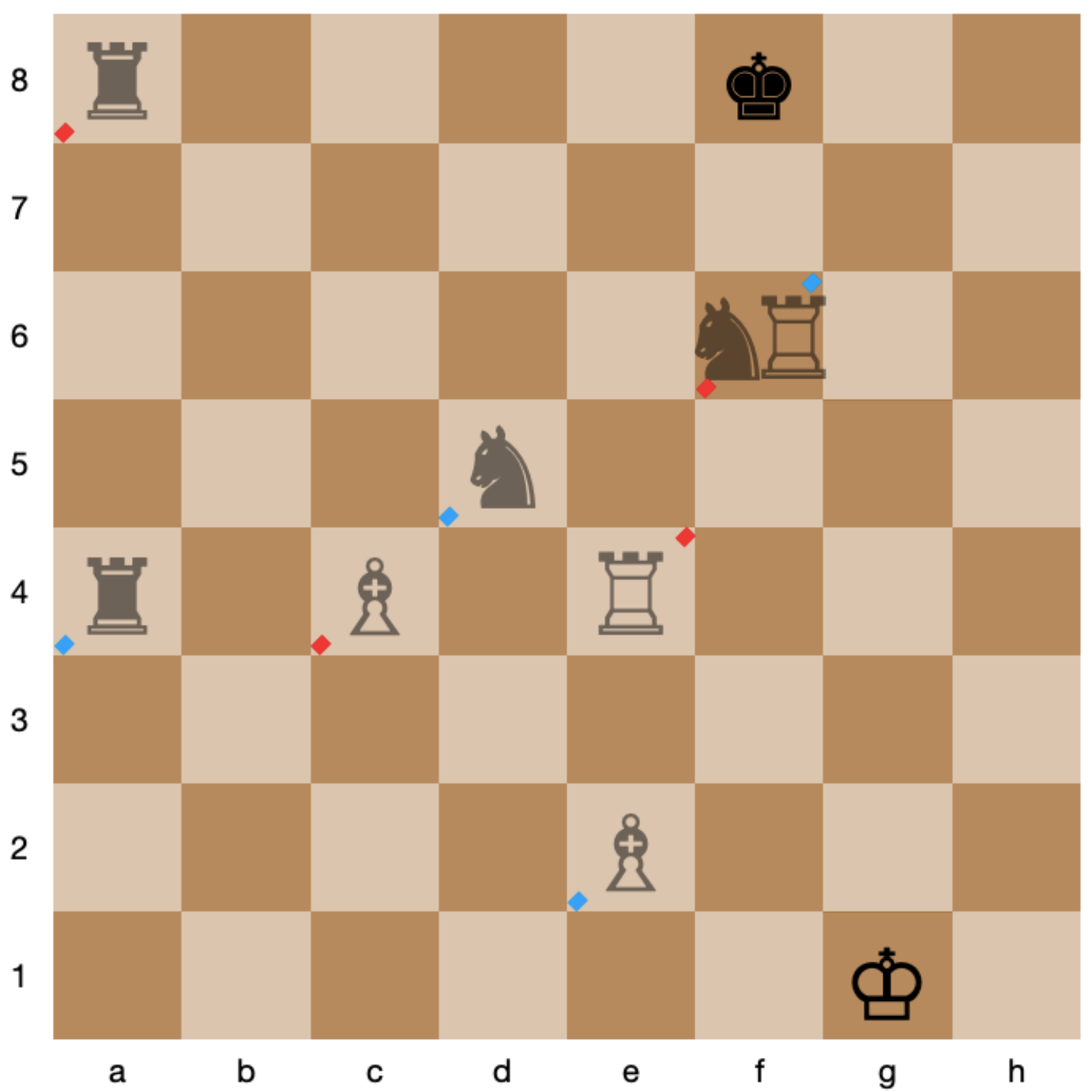}
\vspace*{-0.5cm}
\end{figure}
\begin{figure}[H]
\centering
\includegraphics[width=0.9\columnwidth]{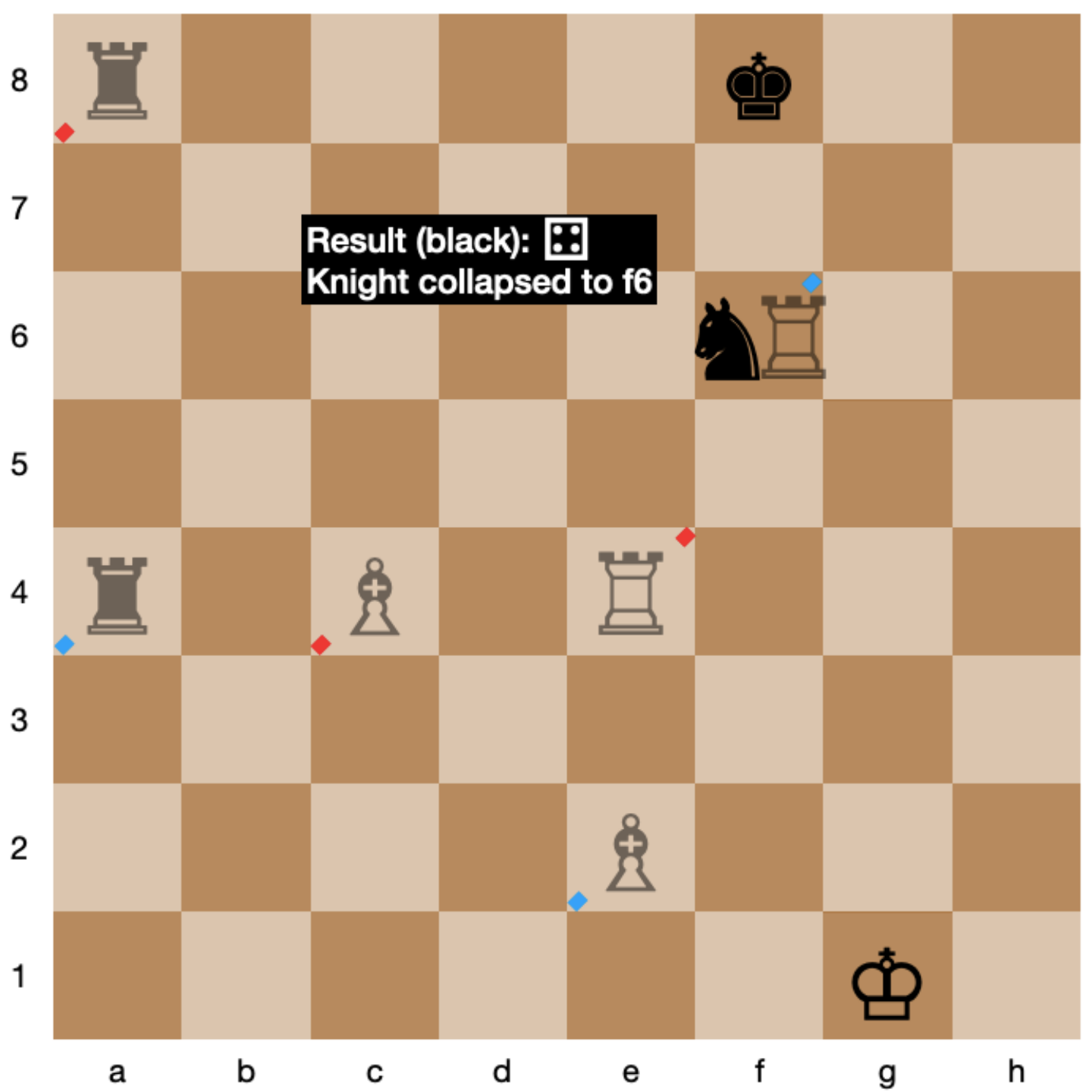}
\vspace*{-0.5cm}
\end{figure}
\begin{figure}[H]
\centering
\includegraphics[width=0.9\columnwidth]{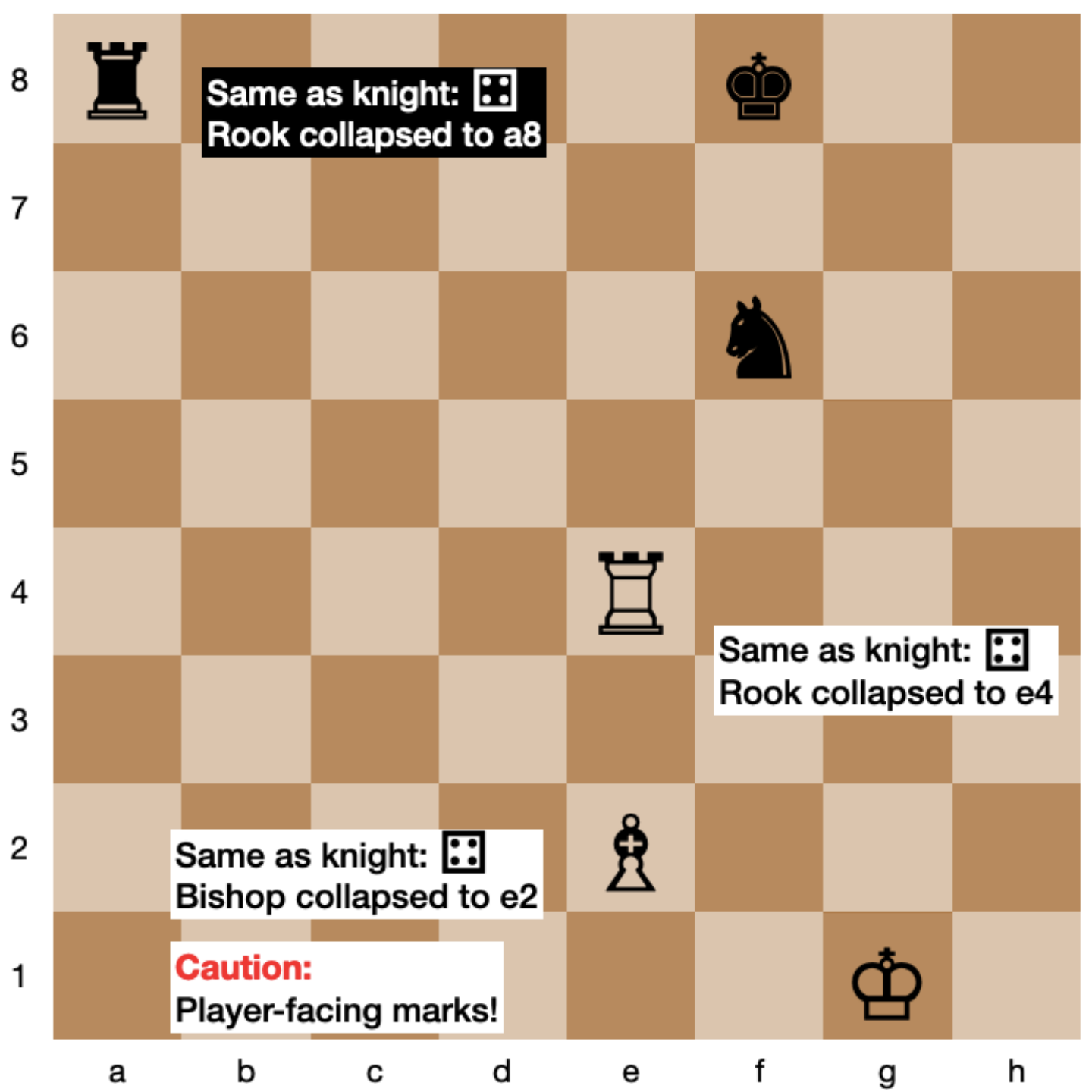}
\end{figure}
\item[8.5] \textit{[Clarification]} In accordance with Articles 3.7.3.1 and 3.7.3.2, the 'en passant' capture of an indefinite pawn which has just advanced two squares in an entanglement first move must be made as though it had moved only one square forward in that entanglement move.\\
\item[8.6] \textit{[Clarification]} The inverting of the facing of the marks as described in Article 7.7 (i.e. rotating the pieces by $180^\circ$) can be performed also with an indefinite pair instance that is entangled with one or more other indefinite pair instances.
\begin{figure}[H]
\centering
\includegraphics[width=0.9\columnwidth]{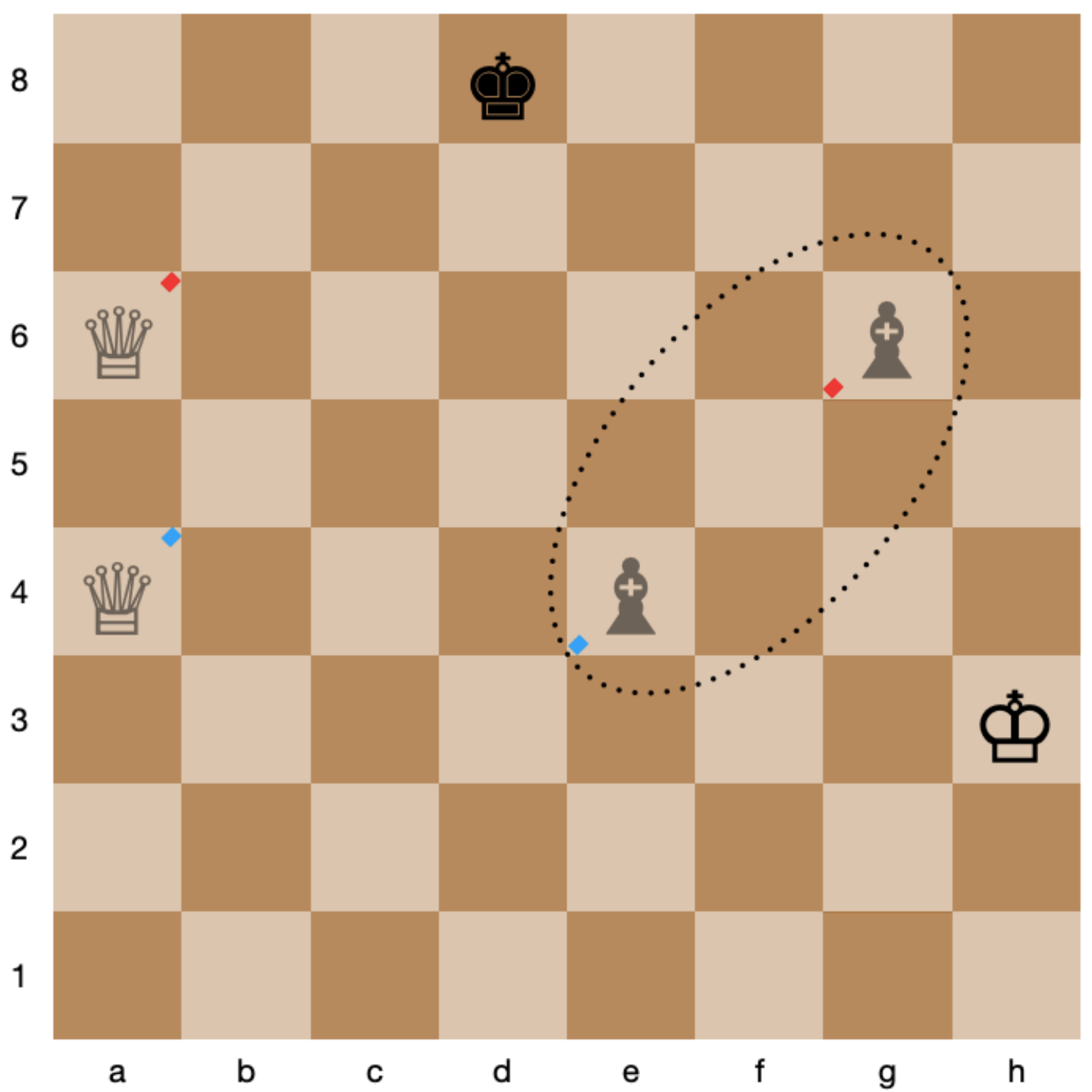}
\vspace*{-0.5cm}
\end{figure}
\begin{figure}[H]
\centering
\includegraphics[width=0.9\columnwidth]{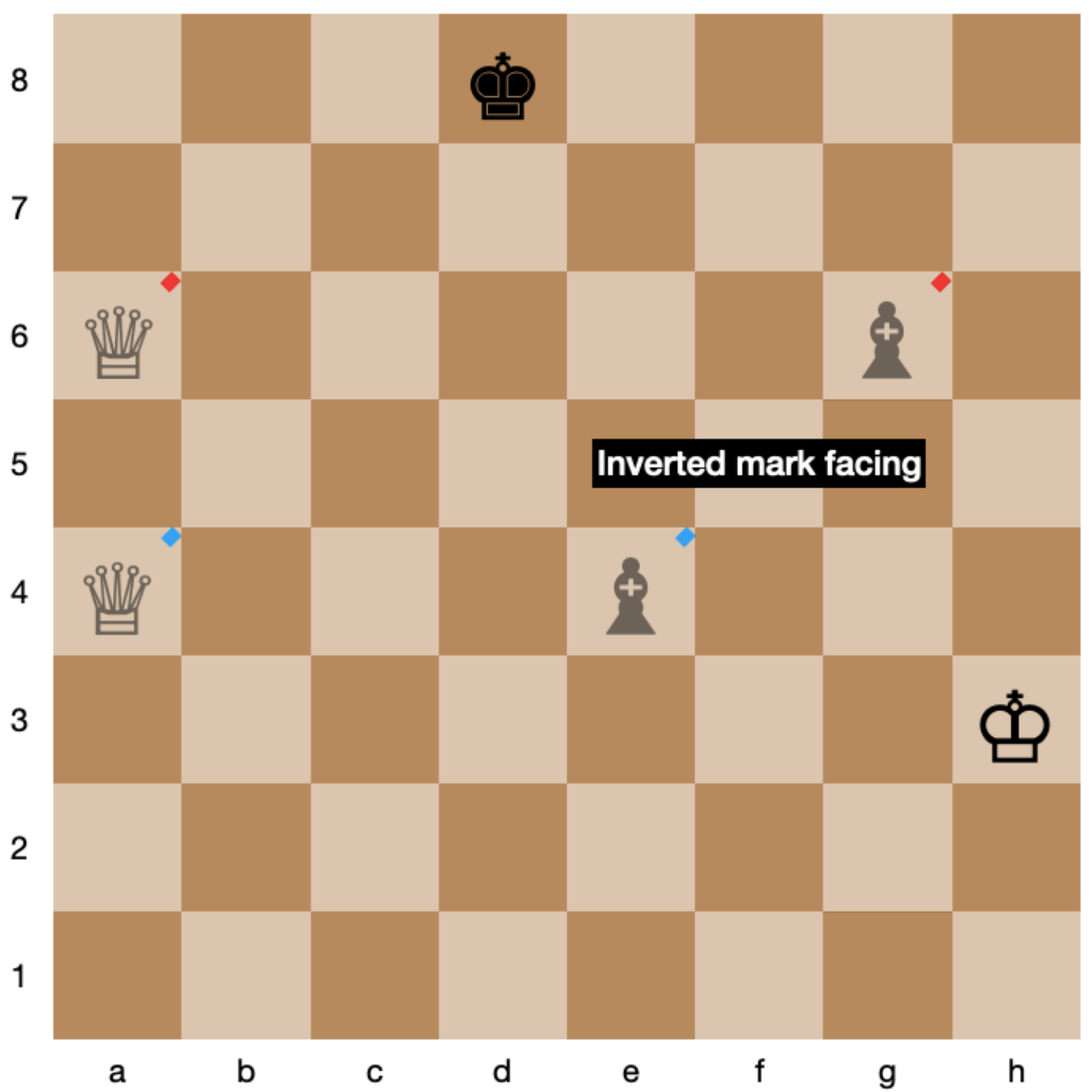}
\end{figure}
\item[8.7] If the king, indefinite or not, is attacked by an opponent's piece, it counts as a 'check' only if a hypothetical capture of that king piece would have a greater than zero chance of success.\\
\item[8.8] The king, indefinite or not, is allowed to be left under, or to be exposed to, attack by an opponent's piece, provided there is zero chance of a successful capture of it.\\
\item[8.9] Attempting to capture the opponent’s king, indefinite or not, is allowed if it has zero chance of success.
\begin{figure}[H]
\centering
\includegraphics[width=0.9\columnwidth]{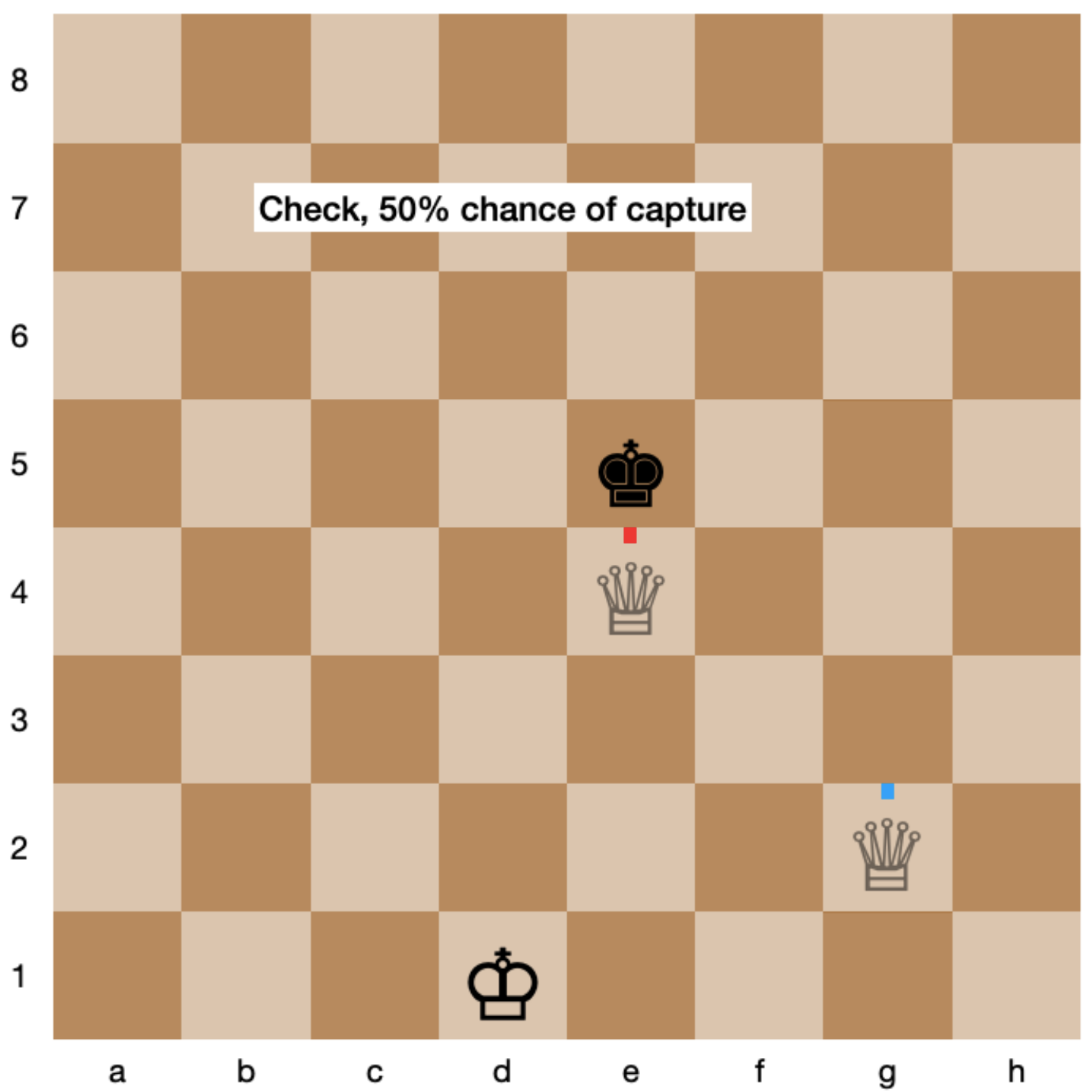}
\vspace*{-0.5cm}
\end{figure}
\begin{figure}[H]
\centering
\includegraphics[width=0.9\columnwidth]{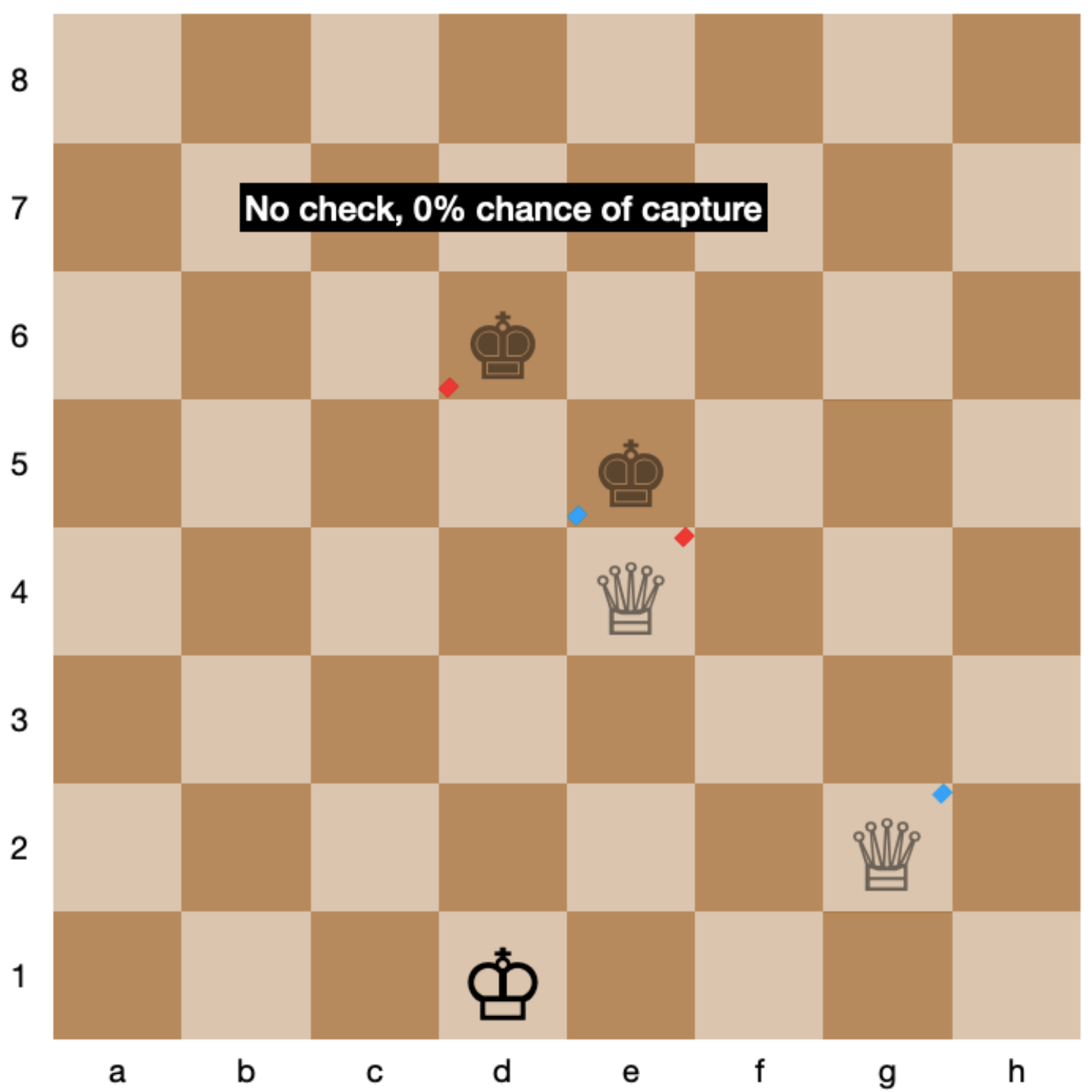}
\end{figure}
\item[8.10] A move is legal when all the relevant requirements of Articles 3.1 to 3.9, Articles 7.1 to 7.9 and Articles 8.1 to 8.9 have been fulfilled. Otherwise, the move is illegal.
\end{enumerate}

\section{Competitive rules of play}\label{competitive}

The detailed specification of competitive rules is out of scope in this paper. A couple of possible adjustments to the basic rules are:

\begin{enumerate}[1.]
\item Dice should be replaced by a quantum random number generator (QRNG) such as  \cite{quantis}, to avoid disputes over random outcomes. QRNG devices generate random numbers based on quantum phenomena, which enables (almost perfect) truly random output that cannot be predicted or influenced by anyone.\\
\item To slow down the game (and likely make it less enjoyable though), the superposition move in Article 7.1 could be modified to only allow that the conventional piece may "move to one unoccupied square and stay where it is at once".\\
\item To somewhat reduce the impact of randomness, it could be allowed to "remotely" collapse an indefinite pair instance owned by the player (incl. those entangled with it), and then immediately make a move with the resulting conventional piece.\\
\item Alternatively, randomness could be completely eliminated, see the 'Deterministic Game' in Appendix E.
\end{enumerate}

\section*{Appendix A: Simplified Game}\label{simplified}

For people who find dealing with the marks overwhelming, an adjustment can be made to the rules that results in a much easier overview of the situation on the chessboard.

The adjustment is the additional Article 9, taking priority over the other articles. The game obeying Article 9 is called the \textit{Simplified Game}. Besides being simpler to play, this game retains both the superposition move and the entanglement move.

\subsubsection*{Article 9: Simplification rule}

\begin{enumerate}[1.1]
\item[9.1] The red mark on an indefinite piece must always face the opponent.\\
\begin{enumerate}[1.1.1]
\item[9.1.1] \textit{[Clarification]} There are two main consequences. First, the inverting of the facing of the marks (see Article 7.7) isn't allowed anymore. Second, in a group of entangled pairs, all pairs will (randomly) collapse to squares with marks of the same colour; that is, the indefinite pieces having the same coloured marks on them are 'connected' within the group.
\end{enumerate}
\end{enumerate}

\section*{Appendix B: The chess set}\label{chessset}

To produce the quantum chess set, there are two options.

One is to have for each conventional piece two corresponding indefinite pieces of the same size, shape and colour, one with a red mark and the other with blue. Ideally, the indefinite pieces are either of a more matte or transparent finish, to be able to better differentiate them. However, they may also be of the same finish as the conventional pieces.

The other option is more economical, where each conventional piece consists of two easily separable halves (held together e.g. by magnetic force), playing the role of the corresponding indefinite pieces. One of them has a red mark on it, the other blue. The marks are inside the conventional piece, pointing laterally outwards, so the act of "facing something" in the quantum rules is represented by "pointing to that thing".

\begin{figure}[H]
\centering
\includegraphics[width=0.9\columnwidth]{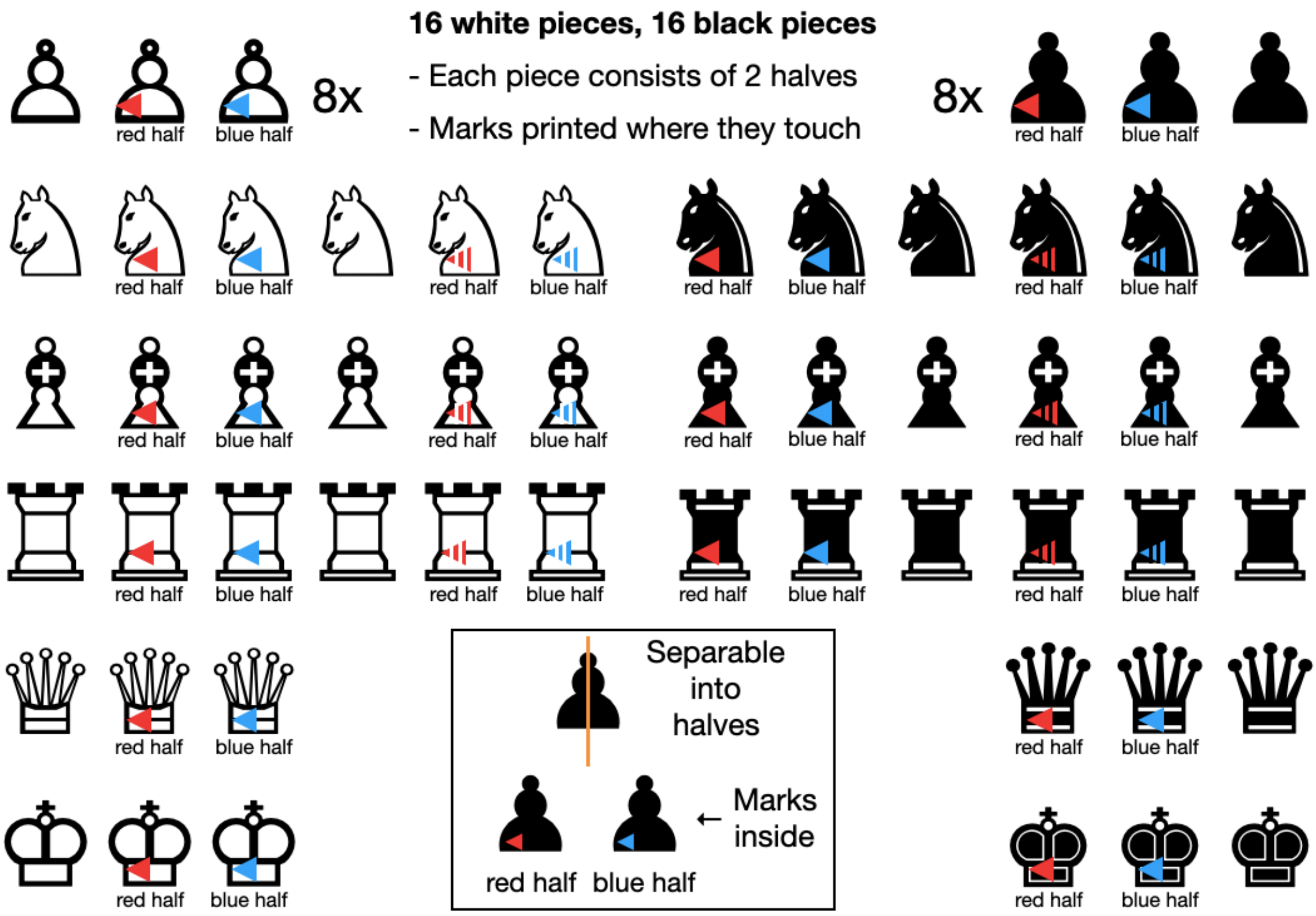}
\end{figure}

The advantage of the first option is that it can be accomplished easily in a DIY manner, by combining (and marking) the pieces of three conventional chess sets.

\section*{Appendix C: Who is Niel?}\label{niel}

\begin{figure}[H]
\centering
\vspace*{-0.5cm}
\includegraphics[width=0.5\columnwidth]{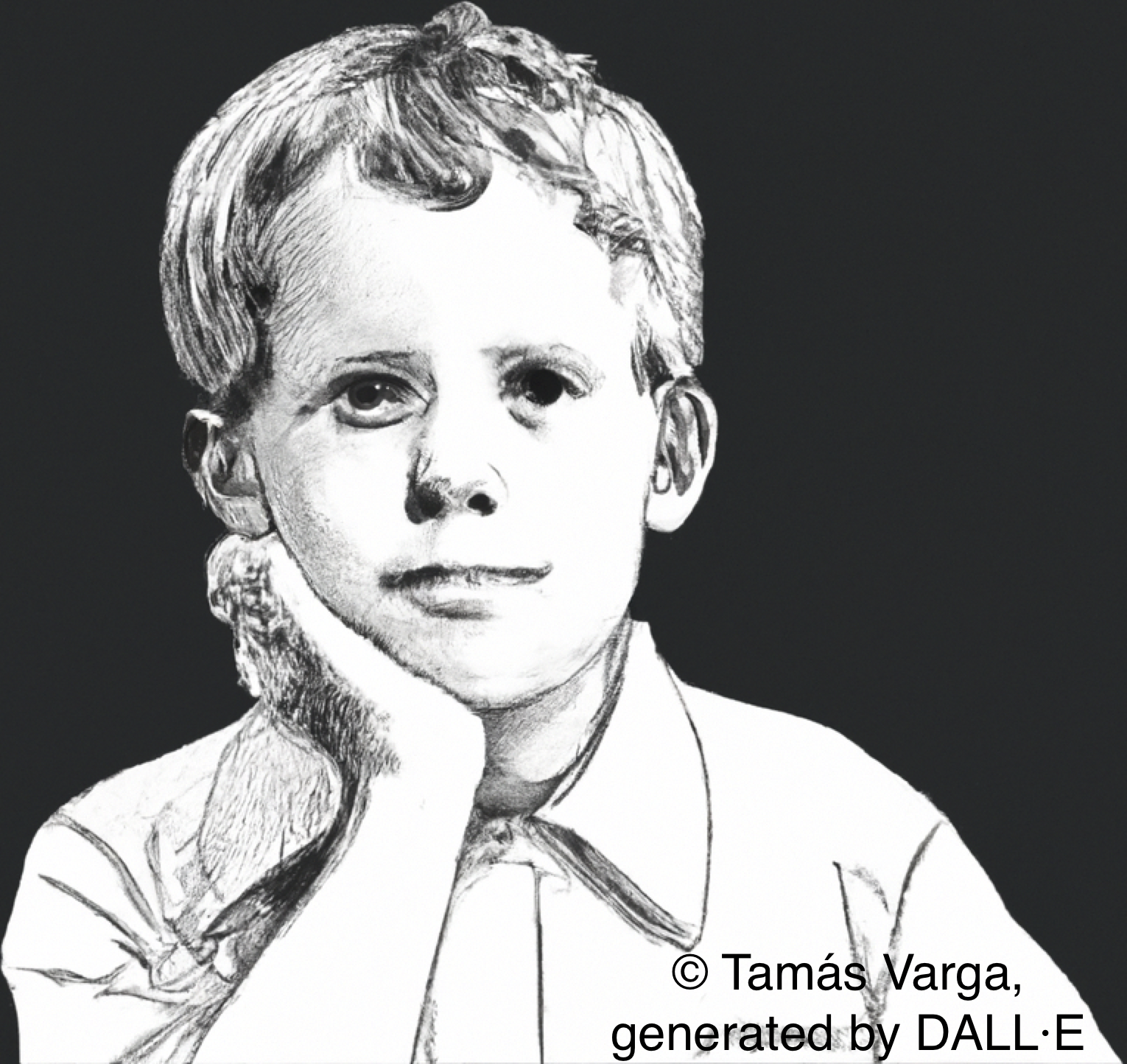}
\end{figure}

\section*{Appendix D: Hadamard Game}\label{hadamard}

To exhibit quantum interference, "unequal superposition" is replaced by "equal superposition with phase". That is, in Article 7.2 it would read: "If one mark faces the opponent and the other the player who owns the pair, it is called an 'equal superposition with phase'."

Then, in step 2 of Article 7.5, the section about unequal superposition is replaced as follows:
\begin{enumerate}[1.]
\item[] - \textbf{Equal superposition with phase: }an even result 2, 4 or 6 (an odd result 1, 3 or 5) means that the player's indefinite pair instance 'collapses' to the square having the indefinite piece on it whose mark faces the opponent (the player).
\end{enumerate}

Finally, Article 7.7 is replaced as well, introducing the \textit{Z move} and the \textit{Hadamard move}:

\begin{enumerate}[1.1]
\item[7.7] The player who owns an indefinite pair may make one of the following moves with that pair, provided that it makes a real difference to the position:\\
\begin{enumerate}[1.1]
\item If one indefinite piece is 'one move away' from the other, the player may invert the facing of the marks on both pieces (by rotating each piece by $180^\circ$).\\
\item \textbf{Z move: }if the pair is not entangled with any other pair, the player may invert the facing of the mark on one of the indefinite pieces.\\
\item \textbf{Hadamard move: }if one indefinite piece is 'one move away' from the other AND the pair is not entangled, the player may replace the pair with a corresponding conventional piece. In case of equal superposition (with phase), the conventional piece must be placed on the square occupied by the indefinite piece with the red mark (the blue mark) on it.\\
\end{enumerate}
\end{enumerate}

\section*{Appendix E: Other variants}\label{variants}

Other variants of Niel's Chess can be devised, by directing and interpreting the marks on the indefinite pieces in different ways.

Even the locations of the indefinite pieces within their respective squares (i.e. up, down, left, right, in the middle, in a corner) might represent a property of a single- or multipartite quantum state. For instance, unequal superposition can be brought back into the Hadamard Game by interpreting the locations of the indefinite pair pieces as follows: "both in the middle" means equal superposition, while "one up, the other down" or "in opposite corners along a diagonal" (for entangled pairs) means unequal superposition.

Also, the Z move and the Hadamard move can be extended to entangled pairs and entangled groups of pairs, respectively (as the latter would involve pieces from both players, maybe the opponent must agree beforehand). This game could be called \textit{Hadamard+ Game}.

Yet another variant would be that in the Hadamard+ Game, an attempted capture would induce Hadamard move(s) instead of random collapse. The rules of check and checkmate would change accordingly. This game could be called \textit{Deterministic Game}.

Niel's Chess can be played on smaller boards as well \cite{minichess}, e.g. on $4\times 5$, $4\times 4$, $4\times 6$  or $5\times 6$ boards with initial positions as follows:

\begin{figure}[H]
\centering
\vspace*{-0.25cm}
\includegraphics[width=0.50\columnwidth]{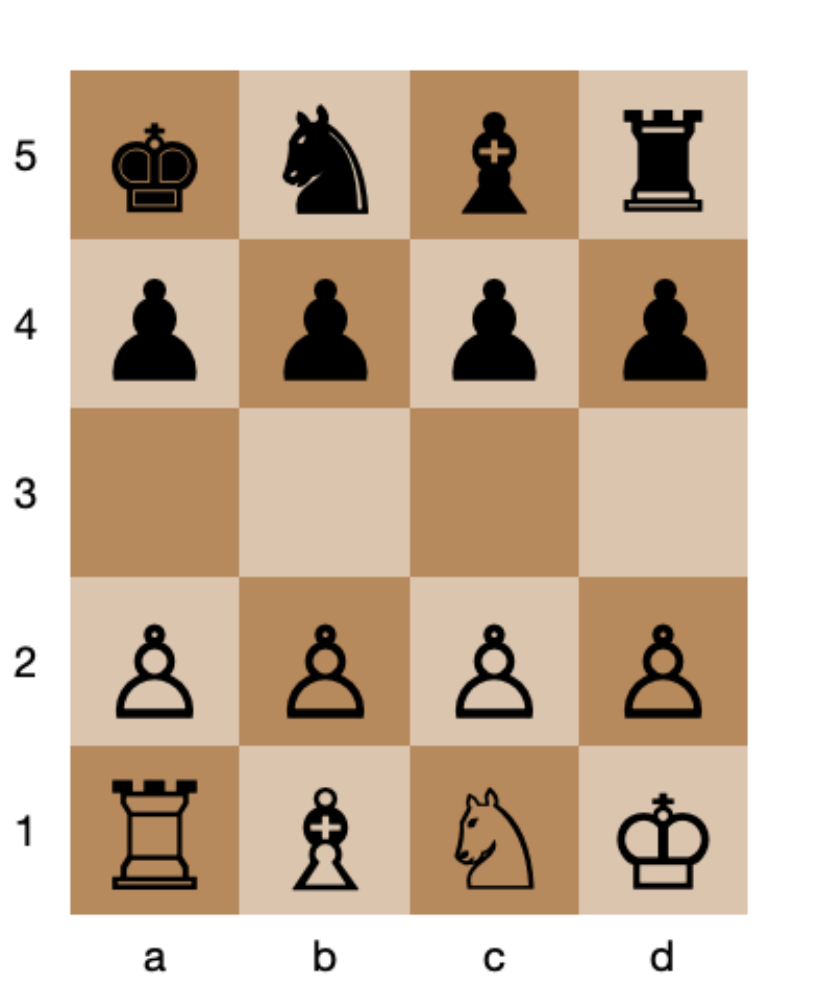}\includegraphics[width=0.50\columnwidth]{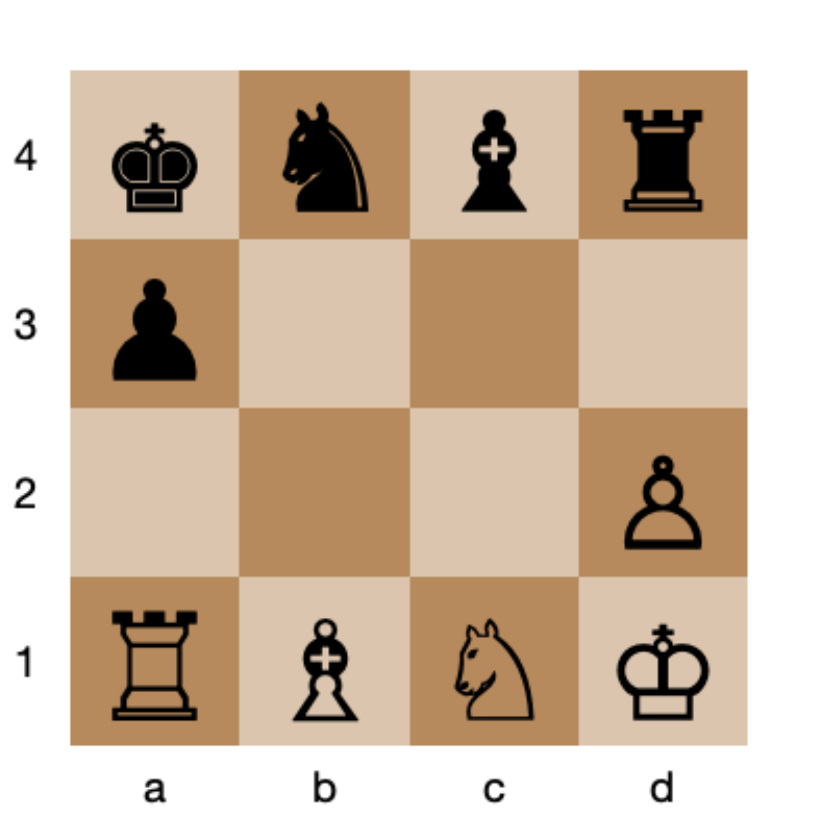}
\vspace*{-1.25cm}
\end{figure}

\begin{figure}[H]
\centering
\includegraphics[width=0.50\columnwidth]{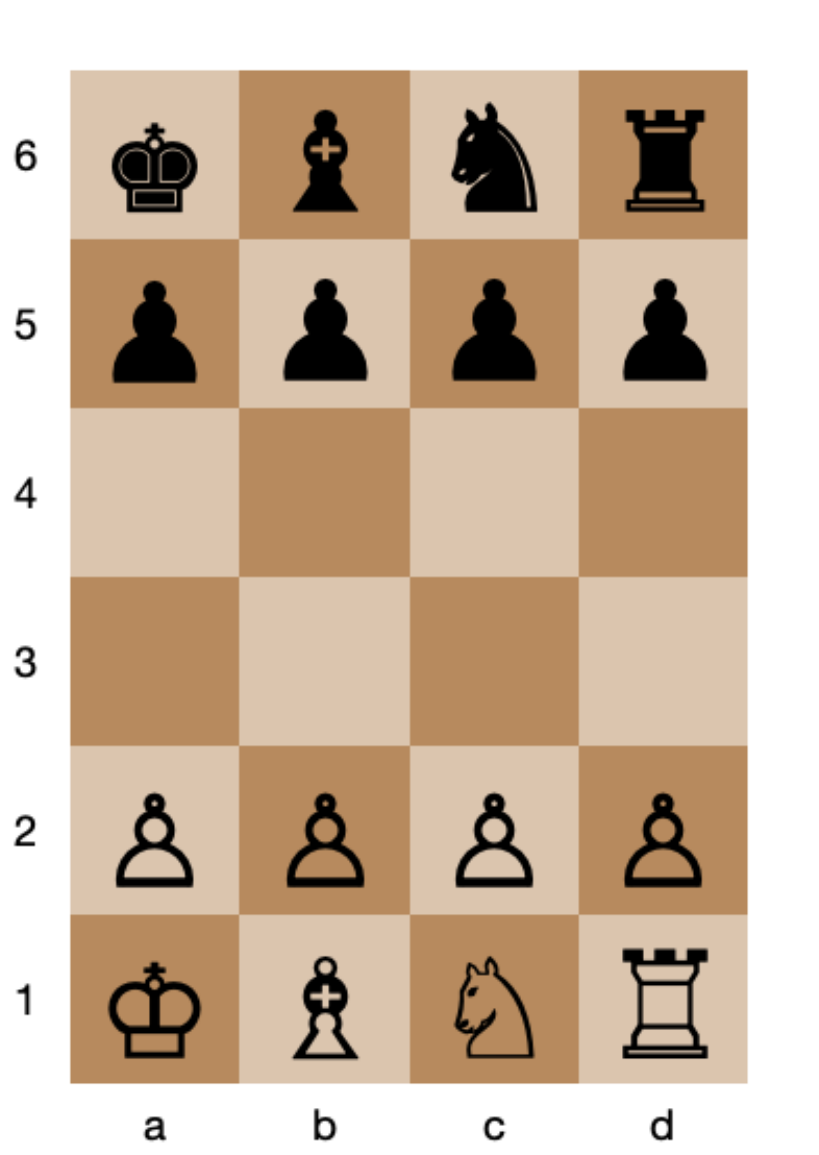}
\vspace*{-1.25cm}
\end{figure}

\begin{figure}[H]
\centering
\includegraphics[width=0.60\columnwidth]{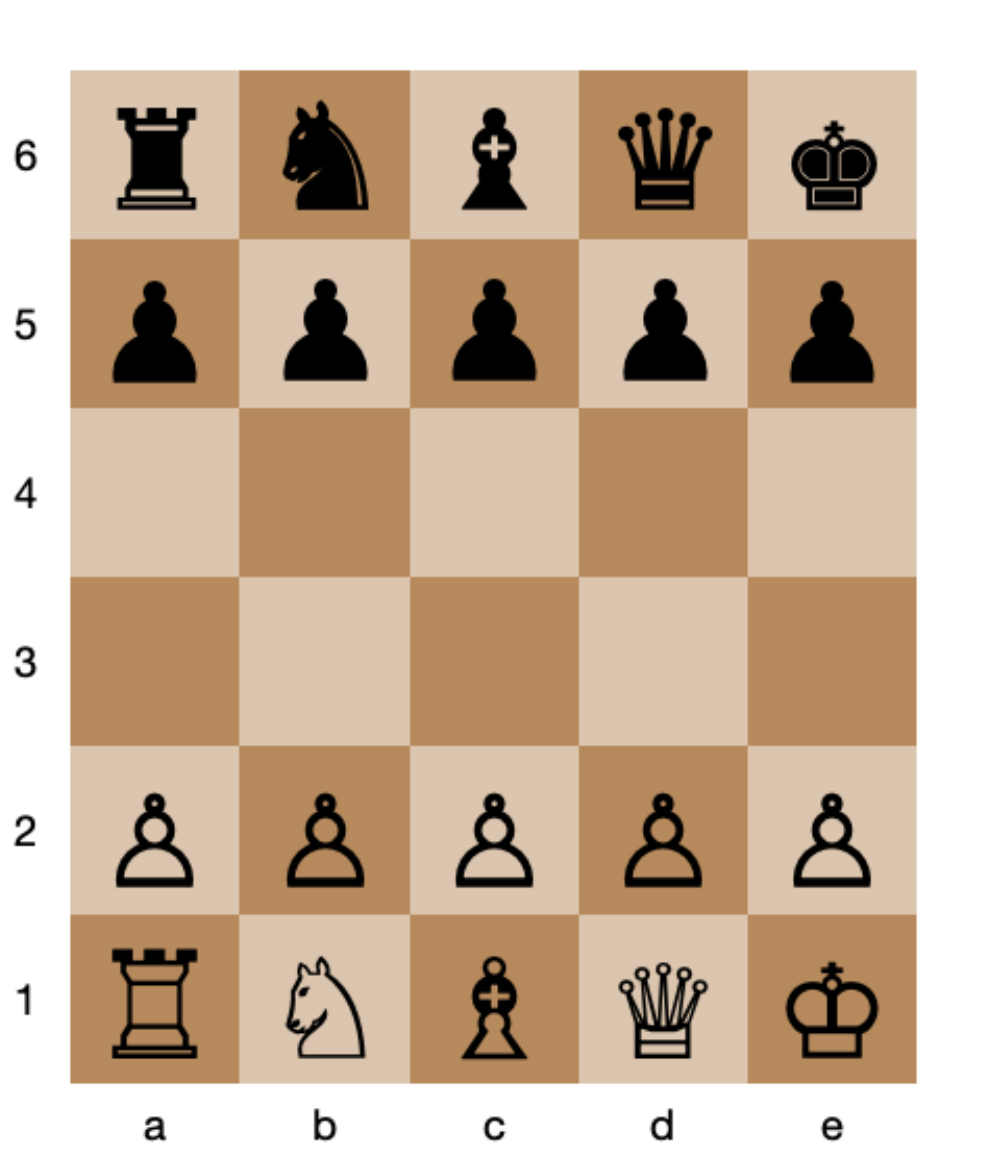}
\vspace*{-0.25cm}
\end{figure}

From an educational perspective, different quantum phenomena can be illustrated by different games (e.g. one could exhibit partial collapse by allowing conventional pieces to be on three or more squares at once, or complex amplitudes via fine-grained rotations, or even quantum tunnelling), and smaller board sizes may be more appropriate for younger players... the limit is the imagination.

An example of tunnelling is shown below: by rolling a 6, the rook successfully tunnelled through an opponent's pawn to an unoccupied square right behind that pawn. Otherwise, by rolling 1 to 5 the rook would have "bounced back" and stayed where it was. Quantum tunnelling may be attempted only when both participating pieces are conventional ones. A game with such a move could be called \textit{Tunnelling Game}.

\begin{figure}[H]
\centering
\includegraphics[width=0.9\columnwidth]{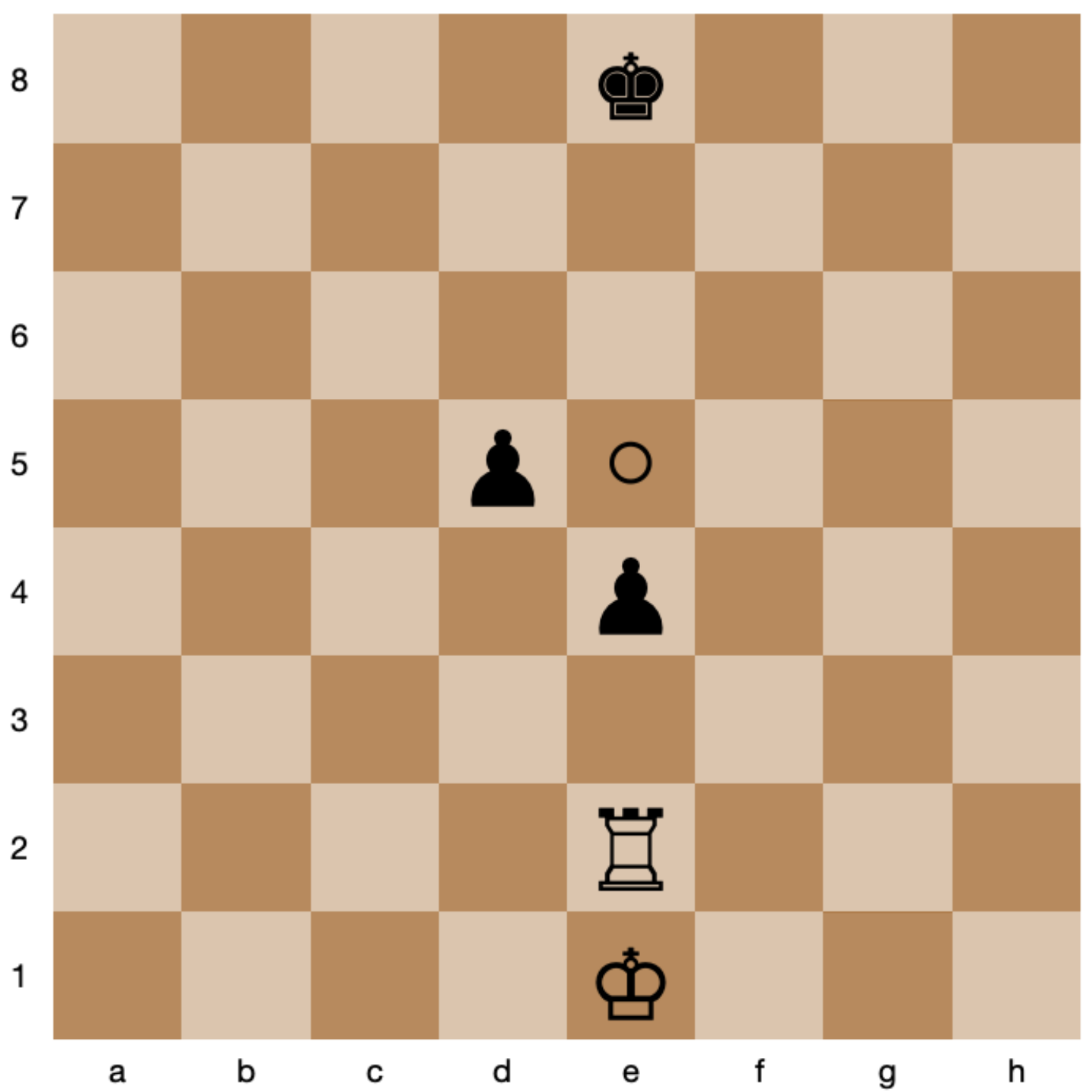}
\vspace*{-0.5cm}
\end{figure}

\begin{figure}[H]
\centering
\includegraphics[width=0.9\columnwidth]{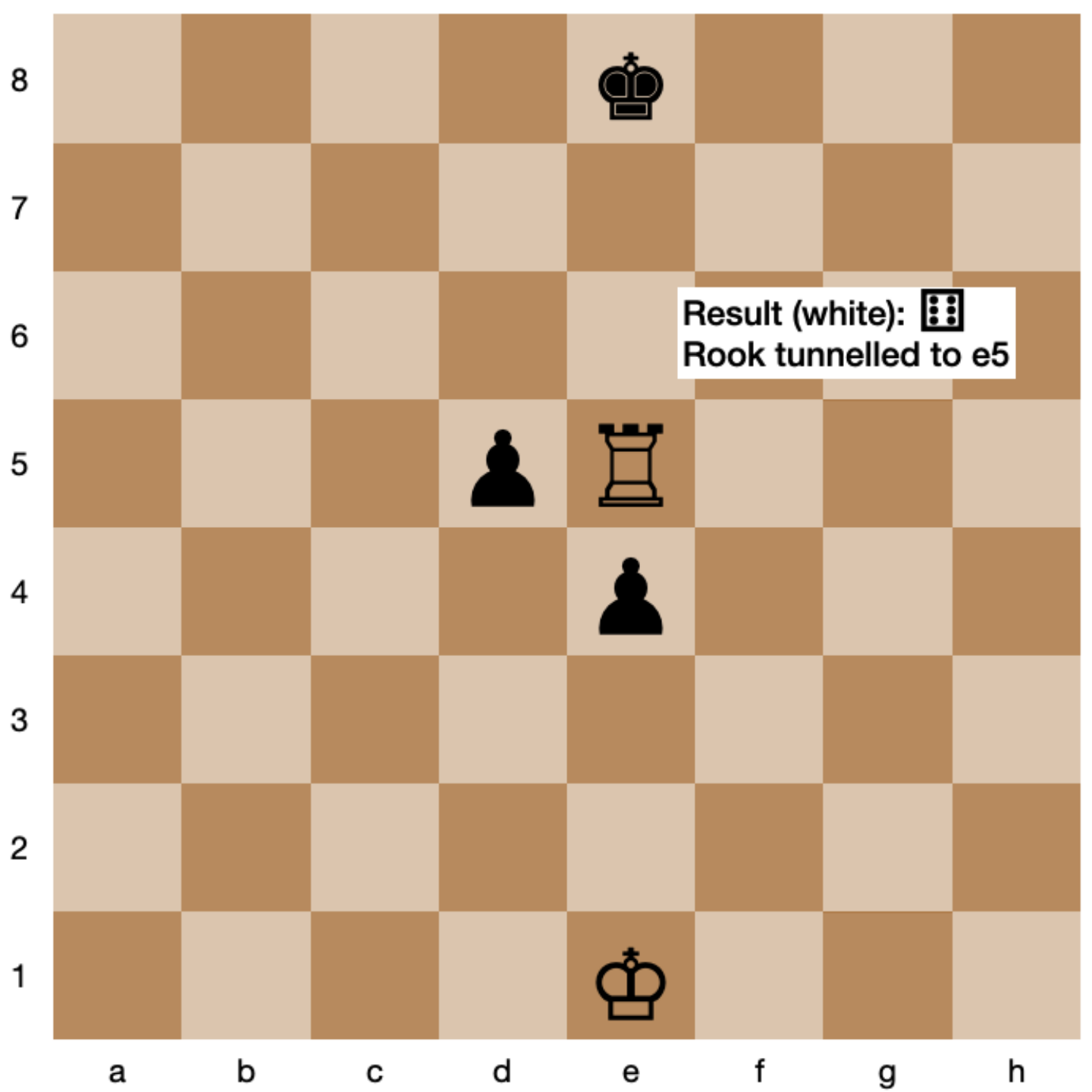}
\vspace*{-0.5cm}
\end{figure}

\begin{figure}[H]
\centering
\includegraphics[width=0.9\columnwidth]{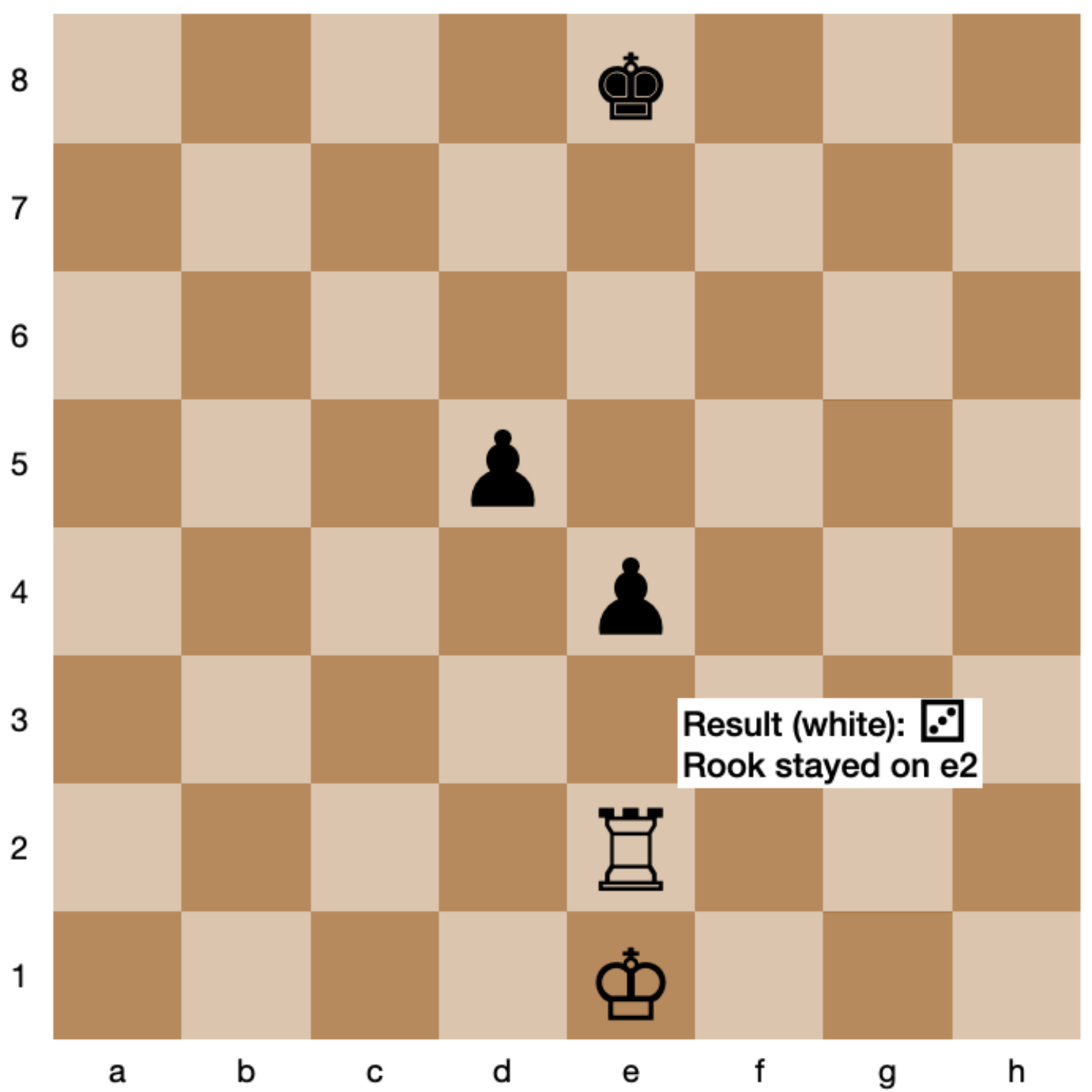}
\end{figure}

\section*{Appendix F: Piece in a box}\label{pieceinbox}

Both Niel's Chess and \cite{cantwell} are based on the spatial superposition of conventional chess pieces. However, there is an important conceptual difference between the two approaches, which can be illuminated in terms of the Schrödinger’s Cat thought experiment \cite{schroedinger}.

The underlying idea in Niel's Chess is “piece in a box”, meaning that each indefinite piece is analogous to half of the sealed box in a Schrödinger’s Cat-like setup. In this setup, instead of being poisoned or not, the cat truly randomly chooses which half of the box to stay in, and then the box is \textit{very carefully} separated into two sealed half-boxes. Later, whenever such a sealed half-box hits, or is hit by, a conventional piece or another sealed half-box, the isolation gets broken and the superposition state inside the whole box collapses immediately.\footnote{For completeness, recall that in \cite{akl} it is the player’s hand touching the piece that causes the (non-spatial) superposition to collapse.}

In contrast, the game in \cite{cantwell} evolves as a single, overall superposition of purely conventional positions on the chessboard. There are no "indefinite pieces", since no piece is isolated individually. It is rather as if the whole chessboard, including the conventional pieces on it, had been placed into a sealed box at the beginning of the game. This idea might be called “chessboard in a box”. In such a setup, no collapse would be necessary per se, because according to quantum theory, informational isolation doesn’t have to be broken in order to move the pieces inside the box via applying unitary operations \cite{schumacher}. Nevertheless, (partial) collapses do happen in the game, targeting one or more squares, triggered by the ”system” due to pragmatic reasons, to keep the complexity of the superposition state inside the box manageable.

The distinct underlying physical viewpoints, i.e. "piece in a box" vs. "chessboard in a box", give rise to characteristic differences between Niel's Chess and \cite{cantwell}, e.g. in the way attempted captures and entanglement moves are executed. These are complemented by several additional differences in gameplay, e.g. the meaning of check and checkmate, unequal superposition and the swapping of collapse probabilities thereof, and that Niel's Chess can be played without relying on a computer. This last feature makes the process of experimenting with new rules easier for educators and players.

\section*{Appendix G: No-box variant}\label{nobox}

An alternative to the piece-in-a-box concept would be to assume that the indefinite pieces don't interact with the environment, so no protecting box is needed.

Then, in Article 7.5, if an indefinite piece attempts to capture an opponent's piece, it would first be collapsed (by the player making the move), and if it doesn't collapse to the intended square of arrival, the attempted capture protocol is aborted, without collapsing the opponent's indefinite piece, if any.

\backmatter

\section*{Declarations}

Niel's Chess is a trademark co-owned by the author, and there is a related patent pending in multiple jurisdictions.


\begin{thebibliography}{9}
\bibitem{mckinsey} McKinsey \& Company, The Rise of Quantum Computing, Web page: \url{https://www.mckinsey.com/featured-insights/the-rise-of-quantum-computing}, accessed 31 March 2023.
\bibitem{akl} S. G. Akl, The Quantum Chess Story, Int. J. Unconv. Comput., 12(2-3), pp. 207-219, 2016.
\bibitem{cantwell} C. Cantwell, Quantum Chess: Developing a Mathematical Framework and Design Methodology for Creating Quantum Games, arXiv: \url{https://arxiv.org/abs/1906.05836}, 2019.
\bibitem{fide} International Chess Federation, FIDE Laws of Chess, Online edition: \url{https://handbook.fide.com/chapter/E012023}, 2022.
\bibitem{quantis} ID Quantique, Quantis QRNG USB, Web page: \url{https://www.idquantique.com/random-number-generation/products/quantis-random-number-generator/}, accessed 31 March 2023.
\bibitem{minichess} Wikipedia, Minichess, Web page: \url{https://en.wikipedia.org/wiki/Minichess}, accessed 31 March 2023.
\bibitem{schroedinger} E. Schrödinger, Die gegenwärtige Situation in der Quantenmechanik, Naturwissenschaften, 23(48), pp. 807-812, 1935.
\bibitem{schumacher} B. Schumacher, M. Westmoreland, \textit{Quantum Processes, Systems, and Information}, Cambridge University Press, Cambridge, 2010.
\end{thebibliography}
\end{document}